\documentstyle[aps,epsf]{revtex} 
\draft
\newcommand{\del}{\partial}

\begin{document}
\title{{\bf Coulomb Effect: A Possible Probe for the Evolution of Hadronic Matter}}
\author{T. Osada\thanks{%
e-mail: osada@fma.if.usp.br} and Y. Hama\thanks{%
e-mail: hama@fma.if.usp.br}}
\address{Instituto de F\'{\i}sica, Universidade de S\~{a}o Paulo, CP 66318,\\
05315-970 S\~{a}o Paulo~-SP, Brazil}
\date{\today }
\maketitle

\begin{abstract}
Electromagnetic field produced in high-energy heavy-ion collisions contains
much useful information, because the field can be directly related to the
motion of the matter in the whole stage of the reaction. One can divide the
total electromagnetic field into three parts, $i.e.$, the contributions from
the incident nuclei, non-participating nucleons and charged fluid, the
latter consisting of strongly interacting hadrons or quarks. Parametrizing
the space-time evolution of the charged fluid based on hydrodynamic model,
we study the development of the electromagnetic field which accompanies the
high-energy heavy-ion collisions. We found that the incident nuclei bring a
rather strong electromagnetic field to the interaction region of hadrons or
quarks over a few fm after the collision. On the other hand, the observed
charged hadrons' spectra are mostly affected (Coulomb effect) by the field
of the charged fluid. We compare the result of our model with experimental
data and found that the model reproduces them well. The pion yield ratio $%
\pi ^{-}/\pi ^{+}$ at a RHIC energy, Au+Au 100+100 GeV/nucleon, is also
predicted.
\end{abstract}


\section{Introduction}

Most of the important informations on the new phase of quark-gluon matter
are obtained by investigating their space-time development in high-energy
heavy-ion collisions. These studies promise not only to specify the
thermodynamical conditions for the opening new phase but also to get useful
knowledge of the evolution of the system. The Bose-Einstein correlations 
\cite{Gyulassy79} attract a great deal of attention as a powerful tool to
study the geometry of the particle emitting sources. However, very
complicated interactions characteristic of these collisions prevent us from
drawing a clear image of the sources.

The Coulomb interaction between a frozen out hadron and the strongly
interacting charged hadronic matter may also be helpful in studying the
evolution of the hadronic matter\cite
{Benenson79,Koonin79,Bertsch80,Gyulassy81,Shuryak91,Wong95,osada96,Barz98} .
One of the advantages of studying its effects is that there is no ambiguity
in constructing the electromagnetic fields if one knows completely the
motion of the hadronic or quark matter. Conversely, if one understands the
electromagnetic fields precisely, one can know the behavior of the hadronic
or quark matter in the high-energy heavy-ion collisions. Another advantage
is that the electromagnetic fields carry information over the whole duration
of the nuclear collision (actually, even of the stage before the collision.)
Fortunately, a great number of charged hadrons are produced in the
high-energy heavy-ion collisions. It is expected that one can obtain
information on the electromagnetic fields by carefully studying their
momentum distributions. The importance of the Coulomb effect at Bevalac
energies has been pointed out by several authors\cite
{Benenson79,Koonin79,Bertsch80,Gyulassy81}. 

The hydrodynamic model \cite{Landau,Bjorken83} describes the space-time
development of the hadronic matter and it can explain various experimental
results. The purpose of this paper is to explore this model a little further
into the problem of the electromagnetic fields due to the expanding charge
distribution produced in high-energy heavy-ion collisions.

In the following, we will confine ourselves only to symmetrical collisions
such as Pb+Pb. The extension to non-symmetrical cases is straightforward. We
will begin by describing in the next section how to compute the number of
participant nucleons and their distribution in the transverse plane. In
order to study the space-time development of the produced electromagnetic
field, we shall assume a longitudinal scaling solution without transverse
expansion for the charged fluid. We find this will be enough to reproduce
the main features of the electromagnetic field in question and moreover, in
computing the ratios of the yields (for instance $\pi ^{-}/\pi ^{+})$,
possible deviations are largely canceled out. In Section I$\!$I$\!$I, we
give account of how to estimate the net charge distribution inside the fluid
along the collision axis. This will be done by using various experimental
results on Pb+Pb 158 GeV/nucleon in the CERN-SPS. Then, in Section I$\!$V,
the electromagnetic field produced by this expanding fluid is computed. In
addition to this, fields by both the incident nuclei and by the penetrating
systems consisting of spectator nucleons are also taken into account%
\footnote{%
In a previous work by Barz $et~al.$ \cite{Barz98}, it is not clear the role
of the incident nuclei and penetrating systems.}. The determination of the
freeze-out hypersurface and computation of the transverse-momentum
distribution is given in Section V. The results obtained are compared with
the experimental data for the pionic yield ratio $\pi ^{-}/\pi ^{+}$ by NA44
Collaboration \cite{NA44-Coulomb}. Finally, we close our discussions with
some concluding remarks and prediction for the pion yield ratio for Au+Au
collisions at 100+100 GeV/nucleon.

\section{Glauber model}

\subsection{Participant and penetrating nucleons in $A$-$B$ collisions}

The participant nucleon number in a nucleus-nucleus collision can be
estimated by using the Glauber model \cite{C-Y-Wong}. To do this, let's
first consider a $h$-$A$ collision. At high energy, one can treat this
problem as a collision of $h$ with a tube in the nucleus (see Fig.1). Here,
we assume that the cross section of the tube is given by the nucleon-nucleon
inelastic cross section $\sigma _{in}\approx 32mb$. The probability that an
inelastic interaction occurs in a volume element $\sigma _{in}\delta z$ is 
\begin{eqnarray}
\delta P(\mbox{\boldmath{$b$}},z_n)=\sigma _{in}\delta z~\rho( %
\mbox{\boldmath{$b$}},z_n)=\sigma _{in}\frac LN\rho (\mbox{\boldmath{$b$}}%
,z_n)\approx \frac{\sigma _{in}}N\int \!\!dz~\rho (\mbox{\boldmath{$b$}},z)
\end{eqnarray}
where $\rho (\mbox{\boldmath{$b$}},z)$ is the nucleon density in the nucleus
and $N$ is the number of equally divided volume elements in the tube (see
Fig.1). The probability that $h$ interacts with $m$ elements in the tube is 
\begin{eqnarray*}
P(m;\mbox{\boldmath{$b$}})= {}_N C_m [\delta P(\mbox{\boldmath{$b$}}%
)]^m[1-\delta P(\mbox{\boldmath{$b$}})]^{N-m}~.
\end{eqnarray*}
Hence, we obtain for the total probability of occurrence of inelastic
reactions in a $h$-$A$ collision 
\begin{eqnarray}
P_{inel}=\sum_{m=1} {}_N C_m [\delta P(\mbox{\boldmath{$b$}})]^m[1-\delta P(%
\mbox{\boldmath{$b$}})]^{N-m}~=1-[~1-\frac{\sigma _{in}}NT(%
\mbox{\boldmath{$b$}})~]^N
\end{eqnarray}
where $T(\mbox{\boldmath{$b$}})$ is the so-called thickness function which
is defined as ~~$\displaystyle{\ T(\mbox{\boldmath{$b$}})\equiv \int
\!\!dz~\rho (\mbox{\boldmath{$b$}},z)}$. In the continuum limit, $%
N\rightarrow \infty $ , this probability can be written as 
\begin{eqnarray}
P_{inel}(\mbox{\boldmath{$b$}})=1-\exp \left[ -\sigma _{in}T(%
\mbox{\boldmath{$b$}})~\right] ~.
\end{eqnarray}

As a natural extension of the $h$-tube collision, one can regard the
nucleus-nucleus collision as a superposition of tube-tube collisions.
Therefore, the total number of participant nucleons and penetrating nucleons
originated from each incident nucleus is given respectively by 
\begin{eqnarray}
&&N_{part}(\mbox{\boldmath{$b$}})=\int \!\!d\mbox{\boldmath{$b$}}^{\prime
}~T(\mbox{\boldmath{$b$}}^{\prime })\left\{ 1-\exp \left[ -\sigma _{in}T(%
\mbox{\boldmath{$b$}}^{\prime }+\mbox{\boldmath{$b$}})~\right] \right\} ~,
\label{N_part} \\
&&N_{pen}(\mbox{\boldmath{$b$}})=\int \!\!d\mbox{\boldmath{$b$}}^{\prime }~T(%
\mbox{\boldmath{$b$}}^{\prime })\exp \left[ -\sigma _{in}T(%
\mbox{\boldmath{$b$}}^{\prime }+\mbox{\boldmath{$b$}})~\right] ~.
\label{N_pene}
\end{eqnarray}
Assuming the nucleon density in the Woods-Saxon form 
\begin{eqnarray*}
\rho (\mbox{\boldmath{$b$}},z)=\frac{\rho _0}{1+\exp \Big[(\sqrt{z^2+%
\mbox{\boldmath{$b$}}^2}-R)/d~\Big]},
\end{eqnarray*}
where $d=0.54$ fm, $R_A=r_0\times A^{1/3}$ and $r_0=1.2$ fm, the penetrating
nucleon number in each hemisphere in Pb+Pb collisions turns out to be 30.7
and 58.4 for the impact parameter $b\leq 3.5$ fm (centrality 5\%) and $b\leq
6.1$ fm (centrality 15\%), respectively. This means that the total number of
participant nucleons in each case is $N^{part}=$353.1 and $N^{part}=$297.7,
respectively. See also Table \ref{table-centrality}. It is gratifying to see
that, when compared with the data (see $N_{B-\bar{B}}$ in the next section),
these numbers are almost exact.

\subsection{Transverse profile of the nucleon number density}

In the present work, we are mostly interested in central collisions. So, we
shall assume a cylindrically symmetrical distribution both of charge and of
fluid density. However, in comparing the results with the data, we have also
to take the experimental conditions into account. It is clear that the
impact parameter is not always zero, but fluctuates from event to event both
in modulus and direction. In order to simplify the computation, we shall
symmetrize the fluid and charge densities by considering this fluctuation by
averaging over $\mbox{\boldmath{$b$}}.$ Furthermore, we assume factorization
of these densities with respect to the longitudinal and transverse variables.

Let us estimate the transverse profile of this symmetrized nucleon number
density. For a given nucleus-nucleus collision, whose geometry is shown in
Fig.2 (with a fixed $\mbox{\boldmath{$b$}}$), and, provided that the {\it %
fluid }contains nucleons which collide {\it inelastically} at least once,
its nucleon density{\it \ }projected into the transverse plane ($%
\mbox{\boldmath{$\rho$}}\equiv (x,y)$) is obtained by adding 
\begin{eqnarray}
T(\mbox{\boldmath{$\rho$}}\mp \frac{\mbox{\boldmath{$b$}}}2)\times \left[
~1-\exp [-\sigma _{in}T(\mbox{\boldmath{$\rho$}}\pm \frac{%
\mbox{\boldmath{$b$}}}2)]~\right] ~.
\end{eqnarray}
So, by averaging over the impact parameter $\mbox{\boldmath{$b$}}$ of the
incident nuclei 
\begin{eqnarray}
T_{fluid}(\mbox{\boldmath{$\rho$}})\!\!=\!\!\frac{\displaystyle{\
2\int_0^{b_{max}}\!\!db~b\int_0^{2\pi }\!\!d\varphi ~T(\mbox{\boldmath{$%
\rho$}}-\frac{\mbox{\boldmath{$b$}}}2)\left[ ~1-\exp [-\sigma _{in}T(%
\mbox{\boldmath{$\rho$}}+\frac{\mbox{\boldmath{$b$}}}2)]~\right] }}{%
\displaystyle{\int_0^{b_{max}}\!\!db~b\int_0^{2\pi }\!\!d\varphi }}~.
\end{eqnarray}
We also have, for the penetrating system and incident nuclei, 
\begin{eqnarray}
T_{pen}(\mbox{\boldmath{$\rho$}})\!\! &=&\!\!\frac{\displaystyle{\
2\int_0^{b_{max}}\!\!db~b\int_0^{2\pi }\!\!d\varphi ~T(\mbox{\boldmath{$%
\rho$}}-\frac{\mbox{\boldmath{$b$}}}2)\left[ ~\exp [-\sigma _{in}T(%
\mbox{\boldmath{$\rho$}}+\frac{\mbox{\boldmath{$b$}}}2)]~\right] }}{%
\displaystyle{\int_0^{b_{max}}\!\!db~b\int_0^{2\pi }\!\!d\varphi }}~, \\
T_{nucl}(\mbox{\boldmath{$\rho$}})\!\! &=&\!\!\frac{\displaystyle{\
2\int_0^{b_{max}}\!\!db~b\int_0^{2\pi }\!\!d\varphi ~T(\mbox{\boldmath{$%
\rho$}}-\frac{\mbox{\boldmath{$b$}}}2)}}{\displaystyle{\int_0^{b_{max}}\!%
\!db~b\int_0^{2\pi }\!\!d\varphi }}~.
\end{eqnarray}

Observe that these quantities are related by 
\begin{equation}
T_{fluid}(\mbox{\boldmath{$\rho$}})+T_{pen}(\mbox{\boldmath{$\rho$}}%
)=T_{nucl}(\mbox{\boldmath{$\rho$}})\,,
\end{equation}
being the integral of the right-hand side equal to $2A$\thinspace . They are
shown in Fig.3. We can also define the normalized densities 
\begin{eqnarray}
\tilde{T}_i(\mbox{\boldmath{$\rho$}})\equiv \frac{\displaystyle{T}_i{(%
\mbox{\boldmath{$\rho$}})}}{\displaystyle{\int \!\!d\rho ~\rho \int
\!\!d\varphi ~{T}_i(\mbox{\boldmath{$\rho$}})}}  \label{renor-thick}
\end{eqnarray}
for the fluid, penetrating system and incident nuclei. In the following
sections, we assume that the charge distribution in each system is
proportional to these quantities. Also, the fluid density will be taken
proportional to $\tilde{T}_{fluid}(\mbox{\boldmath{$\rho$}}).$

\section{Net-charge distribution in the fluid}

\subsection{Space-time evolution of the fluid}

As mentioned in the Introduction, we approximate the evolution of the
charged matter by a longitudinally expanding scaling solution, neglecting
the transverse expansion. The transverse structure has already been
discussed in the previous section. In such a case, the rapidity of the fluid
is given by 
\begin{equation}
{\eta =\frac 12\ln \frac{t+z}{t-z}\;.}  \label{fluid-para}
\end{equation}
It is convenient to take this and the proper time 
\begin{equation}
{\tau =\sqrt{t^2-z^2}\;,}
\end{equation}
together with $\mbox{\boldmath{$\rho$}}$ as the coordinates for describing
the motion of our charged matter. Thus,

\begin{eqnarray}
t=\tau \cosh \eta ,\quad z=\tau \sinh \eta ~,
\end{eqnarray}
for the time and the longitudinal coordinate in the center-of-mass frame.
Strictly speaking, the scaling solution is only valid if one has an
``infinitely long'' fluid ($-\infty \leq \eta \leq +\infty $) with an
infinitely large size and constant density in the transverse directions. For
a real finite fluid, transverse expansion necessarily appears and also the
longitudinal form suffers degradation beginning from the edges. In the
present paper, we will neglect these effects, for the sake of simplicity
because we find that they are not essential as far as the central region of
the collision is concerned.

\subsection{Longitudinal profile of the charge distribution}

Let us now discuss the longitudinal structure of the charge distribution.
The net charge distribution in the $\eta $-variable (integrated charge
distribution over $\mbox{\boldmath{$\rho$}}$-space) in the fluid can be
estimated in terms of the rapidity distributions of the produced particles 
\begin{eqnarray}
\frac{dN_{net}}{dy}\approx \frac{dN_p}{dy}-\frac{dN_{\bar{p}}}{dy}+\frac{%
dN_{\pi ^{+}}}{dy}-\frac{dN_{\pi ^{-}}}{dy}+\frac{dN_{K^{+}}}{dy}-\frac{%
dN_{K^{-}}}{dy}\;.  \label{net_charge}
\end{eqnarray}
It should be noted here that the net charge distribution which we need to
know is the one inside the fluid, so we have to estimate $dN_{net}/dy$ soon
after the freeze-out, $i.e.$, $\tau $=$10^1\sim 10^2$ fm. In this stage
(before the hyperon decays), the multiplicity of the negative mesons ($%
h^{-}\equiv \pi ^{-},K^{-}$) is slightly larger than that of the positive
mesons ($h^{+}\equiv \pi ^{+},K^{+}$) due to the large neutron-proton
asymmetry in large stable nuclei whereas no such Coulomb-energy effect is
expected in the expanding system.

Consider the charge and the baryon number conservations in the collision: 
\begin{eqnarray*}
N_p^{part}\!+N_n^{part}\!\! &=&\!\!\int \!\!dy\frac{dN_{p-\bar{p}}}{dy}%
+\!\int \!\!dy\frac{dN_{n-\bar{n}}}{dy}+\!\int \!\!dy\frac{dN_{Y^{+}-%
\overline{Y^{+}}}}{dy}+\!\int \!\!dy\frac{dN_{Y^0-\overline{Y^0}}}{dy}%
+\!\int \!\!dy\frac{dN_{Y^{-}-\overline{Y^{-}}}}{dy}, \\
N_p^{part}\!\! &=&\!\!\int \!\!dy\frac{dN_{p-\bar{p}}}{dy}+\!\int \!\!dy%
\frac{dN_{{Y^{+}}-\overline{Y^{+}}}}{dy}-\!\int \!\!dy\frac{dN_{{Y^{-}}-%
\overline{Y^{-}}}}{dy}-\!\int \!\!dy\left[ \frac{dN_{h^{-}}}{dy}-\frac{%
dN_{h^{+}}}{dy}\right] _{direct}\!,
\end{eqnarray*}
where $N_p^{part}$ and $N_n^{part}$ are the numbers of participant protons
and neutrons, respectively, and $Y^{+},Y^0$ and $Y^{-}$ stand for positive,
neutral and negative hyperons. For the present, we shall concentrate on
estimating the net charge distribution in the case of 158 GeV/nucleon Pb+Pb
collision for the centrality 5\%. From these two equations, we obtain the
following formula for the negative charge excess in mesons: 
\begin{eqnarray}
\int \!\!\!dy\left[ \frac{dN_{h^{-}}}{dy}-\frac{dN_{h^{+}}}{dy}\right]
_{direct}\hspace*{-8mm}\!\! &=&\!\!N_n^{part}\!-\!\int \!\!dy\frac{dN_{n-%
\bar{n}}}{dy}-\!\int \!\!dy\frac{dN_{\Lambda -\overline{\Lambda }}}{dy}%
-\!\int \!\!\!dy\frac{dN_{\Sigma ^0-\overline{\Sigma ^0}}}{dy}-2\!\int \!\!dy%
\frac{dN_{\Sigma ^{-}-\overline{\Sigma ^{-}}}}{dy}  \nonumber \\
\!\! &\approx &\frac{A-Z}A\times \!N_{B-\bar{B}}-\left\{ ~1.07\times N_{p-%
\bar{p}}+1.6\times N_{\Lambda -\overline{\Lambda }}~\right\} ~.
\end{eqnarray}
Here, we used the experimental results, $N_{n-\overline{n}}/N_{p-\bar{p}%
}\!\approx $1.07, $N_{\Sigma ^{\pm }-\overline{\Sigma ^{\pm }}}/N_{\Lambda -%
\bar{\Lambda}}\approx $ 0.3 \cite{NA49-Roland,NA49-Appelshauser}. In the
last term above, $N_{\Lambda -\overline{\Lambda }}$ contains also $\Sigma ^0-%
\overline{\Sigma ^0}.$ One can then estimate the negative charge excess
using the experimental value for the net baryon number $N_{B-\bar{B}}\,$=352$%
\pm $12, being the net proton number $N_{p-\bar{p}}$\thinspace = 148.9 (See
Fig.4) and $N_{\Lambda -\overline{\Lambda }}\,$=27.4 \cite{NA49-Appelshauser}%
, as 
\begin{eqnarray}
N_{h^{-}-h^{+}}\!\!\equiv \!\!\int \!\!dy\left[ \frac{dN_{h^{-}}}{dy}-\frac{%
dN_{h^{+}}}{dy}\right] _{direct}\hspace*{-5mm}\approx 10.0.
\label{neg_excess}
\end{eqnarray}
One can verify the consistency of this estimate, by computing the
participant proton number $N_p^{part}$ which is approximately $Z/A\times
N_{B-\bar{B}}\approx $ 138.7 and comparing with the total net charge,
obtained by using $N_{p-\bar{p}}$ data and eq. (\ref{neg_excess}): $N_{p-%
\bar{p}}-N_{h^{-}-h^{+}}\approx $ 138.9.

Hence, by putting eq. (\ref{neg_excess}) into eq. (\ref{net_charge}) and by
considering that the existing negative particle data include also $\bar{p}%
\,, $ finally we arrive at the following expression for the charge
distribution of particles at the freeze-out time 
\begin{eqnarray}
\frac{dN_{net}}{dy}\!\! &\approx &\!\!\frac{dN_p}{dy}-\frac{dN_{\bar{p}}}{dy}%
+\left\{ \frac{dN_{\pi ^{+}}}{dy}+\frac{dN_{K^{+}}}{dy}\right\} -\left\{ 
\frac{dN_{\pi ^{-}}}{dy}+\frac{dN_{K^{-}}}{dy}\right\}  \nonumber \\
\!\! &\approx &\!\!\frac{dN_{p-\bar{p}}}{dy}-\frac{N_{h^{-}-h^{+}}}{\langle
N_{h^{-}+\bar{p}}\rangle -\langle N_{\bar{p}}\rangle }\times \left\{ \frac{%
dN_{h^{-}+\bar{p}}}{dy}-\frac{dN_{\bar{p}}}{dy}\right\} \;.
\label{eq-netcharge}
\end{eqnarray}
For Pb+Pb at 158 GeV/nucleon, we have $N_{h^{-}+\bar{p}}\,$= 695$\pm $30~%
\cite{NA49-Appelshauser} (see Fig.5). The $N_{\bar{p}}$ is estimated by
using an empirical relation $\bar{p}/\Lambda $= 0.12$\sim $0.13 found in $%
S+S $ and $S+Ag$ collisions at 200GeV/nucleon \cite{NA35-1,NA35-2}. Because
the multiplicity of $\bar{\Lambda}\approx $7.5 ($\bar{\Lambda}/\Lambda
\approx $0.2) \cite{NA49-Bormann}, we obtain $N_\Lambda \approx $ 35 and $N_{%
\bar{p}}\approx $ 4.4 (See Fig.6).~The solid line in Fig.7 shows the net
charge distribution obtained by eq.(\ref{eq-netcharge}), for the centrality
5\%, in 158 GeV/nucleon Pb+Pb.

Now, the rapidity distribution of produced particles given by eq.(\ref
{eq-netcharge}) is a convolution of the charge rapidity distribution in the
fluid and the thermal rapidity distribution of the produced particles 
\begin{eqnarray}
\frac{dN_{net}}{dy}\simeq {\cal {N}}\int \!\!d\eta \frac{dN_{net}}{d\eta }%
\times \frac 1{e^{\langle m_{\mbox{\tiny T}}\rangle \cosh (y-\eta )/T}-1},
\label{eq:net-fluid}
\end{eqnarray}
where $T=140$ MeV is the freeze-out temperature and {\bf $\langle m_{%
\mbox{\tiny T}}\rangle $ }is the pion mean transverse mass, evaluated with
the Bose distribution at the temperature{\bf \ $T$.} So, in computing the
electromagnetic field in the following section, we took a piece-wise flat
rapidity distribution for the fluid net charge as seen in Fig.7 which, upon
convolution eq.(\ref{eq:net-fluid}), approximates as closely as possible the
particle distribution of eq.(\ref{eq-netcharge}). The normalization constant 
${\cal {N}}$ in eq.(\ref{eq:net-fluid}) is fixed by equating the integrals
of $dN_{net}/dy$ and $dN_{net}/d\eta $. As for the fluid density, as
mentioned in the beginning of this section we took a completely flat $\eta -$%
distribution which gives the correct number of pions.

\subsection{Centrality dependence of the net charge distribution}

As is easily understood intuitively, the net charge distribution strongly
depends on the centrality of the collision. In the previous subsection, we
have estimated the net charge distribution by using the NA49 data on 158
GeV/nucleon Pb+Pb collision with the centrality 5 \%\thinspace \cite
{NA49-Appelshauser}. We would like to study the Coulomb effect on the $\pi
^{+}/\pi ^{-}$ ratio, which has been measured by the NA44 Collaboration \cite
{NA44-Coulomb}, considering events with 15 \% centrality. The question is
that there are no corresponding data which would allow us to estimate the
net charge distribution under the same experimental conditions. We have
mentioned in Section I$\!$I and verified in Section I$\!$I$\!$I that the
Glauber model we described there gives a very good account of the NA49 data.
In view of this, we will account for the centrality dependence of the net
charge distribution, by the use of the referred model. Concretely, by
assuming that the longitudinal structure of the net charge distribution
remains the same, changing only the transverse profile and consequently the
normalization, we introduce the following factor to relating quantities with
different centrality ($\chi $):

\begin{eqnarray*}
\frac{dN_h}{d\eta }\bigg|_{centrality~\chi ^{\prime }}=\frac{G(\chi ^{\prime
})}{G(\chi )}\frac{dN_h}{d\eta }\bigg|_{centrality~\chi }\,\;,
\end{eqnarray*}
for $h=h^{-}\!\!+\bar{p}$ and $p-\bar{p}\,,$ where $G(\chi )=N^{part}$ for
the centrality $\chi \,$. The participant nucleon-number dependence of
hyperon multiplicity is studied in Ref.\cite{Andersen98}. Using their
results, we have 
\begin{eqnarray*}
\hspace*{15mm}\frac{dN_Y}{d\eta }\bigg|_{centrality~\chi ^{\prime }}=\left[ ~%
\frac{G(\chi ^{\prime })}{G(\chi )}~\right] ^{1.30}\frac{dN_Y}{d\eta }\bigg|%
_{centrality~\chi }\;,
\end{eqnarray*}
for $Y$=$\Lambda $ hyperon. The multiplicity values estimated for the
centrality 15\thinspace \%\thinspace , by using the data obtained for the
centrality 5\thinspace \%\thinspace , are summarized in Table \ref
{table-centrality}.

\section{Electromagnetic\ fields\ in\ high-energy\ heavy-ion\ collisions}

Having been defined the charge distributions and their time development in
the previous sections, now we can compute the electromagnetic field produced
by each one of such distributions.

\subsection{Potentials of the longitudinal expanding charged matter}

An infinitesimal charge $dQ$ moving along a trajectory $\mbox{\boldmath{$r$}}%
=\mbox{\boldmath{$r$}}_0(t)$ in the hadronic or quark matter (fluid)
produces a Li\'{e}nard-Wiechert four-vector potential 
\begin{eqnarray}
A^\mu (t,\mbox{\boldmath{$x$}})=\left( \!\frac{dQ}{|\mbox{\boldmath{$x$}}-%
\mbox{\boldmath{$r$}}_0(t^{\prime })|-\mbox{\boldmath{$v$}}(t^{\prime
})\cdot \{\mbox{\boldmath{$x$}}-\mbox{\boldmath{$r$}}_0(t^{\prime })\}}~,~%
\frac{\mbox{\boldmath{$v$}}dQ}{|\mbox{\boldmath{$x$}}-\mbox{\boldmath{$r$}}%
_0(t^{\prime })|-\mbox{\boldmath{$v$}}(t^{\prime })\cdot \{%
\mbox{\boldmath{$x$}}-\mbox{\boldmath{$r$}}_0(t^{\prime })\}}\!\right) ~,
\label{A-filed}
\end{eqnarray}
where $t^{\prime }$ is the retarded time satisfying the condition, 
\begin{eqnarray}
t^{\prime }+|\mbox{\boldmath{$x$}}-\mbox{\boldmath{$r$}}_0(t^{\prime })|=t~~.
\label{time-condition}
\end{eqnarray}
In terms of the coordinates $(\tau ,\mbox{\boldmath{$\rho$}},\eta ),$ for $%
\mbox{\boldmath{$r$}}_0(t^{\prime })=(\mbox{\boldmath{$\rho$}}^{\prime
},~v_zt^{\prime })\equiv (\mbox{\boldmath{$\rho$}}^{\prime },~\tau ^{\prime
}\sinh \eta ^{\prime })$, eqs.(\ref{A-filed}) and (\ref{time-condition}) may
be rewritten 
\begin{eqnarray}
A^\mu (\tau ,\mbox{\boldmath{$\rho$}},\eta )=\left( ~\frac{\cosh \eta
^{\prime }~dQ}{\tau \cosh (\eta -\eta ^{\prime })-\tau ^{\prime }}~,0,~0,~%
\frac{\sinh \eta ^{\prime }~dQ}{\tau \cosh (\eta -\eta ^{\prime })-\tau
^{\prime }}~\right) ,
\end{eqnarray}
where 
\begin{eqnarray}
\tau ^{\prime }{}^2-2\tau \left[ \cosh (\eta -\eta ^{\prime })\right] \tau
^{\prime }+\tau ^2-|\mbox{\boldmath{$\rho$}}-\mbox{\boldmath{$\rho$}}%
^{\prime }|^2=0~.  \label{time-condition2}
\end{eqnarray}
By solving eq.(\ref{time-condition2}) with the condition $\tau \ge \tau
^{\prime }$, we have 
\begin{eqnarray}
\tau ^{\prime }=\tau \cosh (\eta -\eta ^{\prime })-\sqrt{\tau ^2\sinh
^2(\eta -\eta ^{\prime })+|\mbox{\boldmath{$\rho$}}-\mbox{\boldmath{$\rho$}}%
^{\prime }|^2}~  \label{taup}
\end{eqnarray}
with $\tau \ge |\mbox{\boldmath{$\rho$}}-\mbox{\boldmath{$\rho$}}^{\prime }|$%
~.

As discussed in the previous sections, the charge distribution in the
expanding fluid is written as 
\begin{eqnarray}
\frac{dQ(\eta ^{\prime },\mbox{\boldmath{$\rho$}}^{\prime })}{d%
\mbox{\boldmath{$\rho$}}^{\prime }d\eta ^{\prime }}=e~\tilde{T}_{fluid}(%
\mbox{\boldmath{$\rho$}}^{\prime })~\frac{dN_{net}}{d\eta ^{\prime }},
\end{eqnarray}
so the total potentials due to the whole fluid, $A^0\equiv \phi
_{fluid}(\tau ,\mbox{\boldmath{$\rho$}},\eta )$ and $A^3\equiv
A_{fluid}(\tau ,\mbox{\boldmath{$\rho$}},\eta ),$ are given by 
\begin{equation}
\phi _{fluid}(\tau ,\mbox{\boldmath{$\rho$}},\eta )\!\!=e\int \!\!d%
\mbox{\boldmath{$\rho$}}^{\prime }~\tilde{T}_{fluid}(\mbox{\boldmath{$\rho$}}%
^{\prime })~\int \!\!d\eta ^{\prime }~\frac{dN_{net}}{d\eta ^{\prime }}\frac{%
\cosh \eta ^{\prime }\times \theta (\tau -|\mbox{\boldmath{$\rho$}}-%
\mbox{\boldmath{$\rho$}}^{\prime }|)}{\sqrt{\tau ^2\sinh ^2(\eta -\eta
^{\prime })+|\mbox{\boldmath{$\rho$}}-\mbox{\boldmath{$\rho$}}^{\prime }|^2}}
\end{equation}
and 
\begin{equation}
A_{fluid}(\tau ,\mbox{\boldmath{$\rho$}},\eta )\!\!=\!\!\;e\int \!\!d%
\mbox{\boldmath{$\rho$}}^{\prime }~\tilde{T}_{fluid}(\mbox{\boldmath{$\rho$}}%
^{\prime })~\int \!\!d\eta ^{\prime }~\frac{dN_{net}}{d\eta ^{\prime }}\frac{%
\sinh \eta ^{\prime }\times \theta (\tau -|\mbox{\boldmath{$\rho$}}-%
\mbox{\boldmath{$\rho$}}^{\prime }|)}{\sqrt{\tau ^2\sinh ^2(\eta -\eta
^{\prime })+|\mbox{\boldmath{$\rho$}}-\mbox{\boldmath{$\rho$}}^{\prime }|^2}}%
~,
\end{equation}
respectively.

\subsection{Potentials of the incident nuclei and the penetrating nuclear
systems}

The incident nuclei ({\it nucl}) and the penetrating nuclear systems which
are not included in the fluid ({\it pen}) also produce electromagnetic
fields. Here, we assume that the penetrating systems consist of nucleons
which suffered no inelastic collision and take their rapidities $y$ the same
as the beam rapidities $\pm \,y_{be}$. The scalar and vector potentials can
be obtained in a similar way as in the case of fluid. The potentials due to
the incident nuclei are 
\begin{equation}
\phi _{nucl}(\tau ,\mbox{\boldmath{$\rho$}},\eta )\!\!=\,\!\!eZ\!\int \!\!d%
\mbox{\boldmath{$\rho$}}^{\prime }~\tilde{T}_{nucl}(\mbox{\boldmath{$\rho$}}%
^{\prime })\sum_{\eta ^{\prime }=\pm ~y_{be}}\left[ ~\frac{\cosh \eta
^{\prime }\times \theta (|\mbox{\boldmath{$\rho$}}-\mbox{\boldmath{$\rho$}}%
^{\prime }|-\tau )}{\sqrt{\tau ^2\sinh ^2(\eta -\eta ^{\prime })+|%
\mbox{\boldmath{$\rho$}}-\mbox{\boldmath{$\rho$}}^{\prime }|^2}}~\right] \ 
\label{scl-nucl}
\end{equation}
and 
\begin{equation}
A_{nucl}(\tau ,\mbox{\boldmath{$\rho$}},\eta )\!\!=\!\!\;eZ\!\int \!\!d%
\mbox{\boldmath{$\rho$}}^{\prime }~\tilde{T}_{nucl}(\mbox{\boldmath{$\rho$}}%
^{\prime })\sum_{\eta ^{\prime }=\pm ~y_{be}}\left[ ~\frac{\sinh \eta
^{\prime }\times \theta (|\mbox{\boldmath{$\rho$}}-\mbox{\boldmath{$\rho$}}%
^{\prime }|-\tau )}{\sqrt{\tau ^2\sinh ^2(\eta -\eta ^{\prime })+|%
\mbox{\boldmath{$\rho$}}-\mbox{\boldmath{$\rho$}}^{\prime }|^2}}~\right] ~,
\label{vec-nucl}
\end{equation}
where $Z$ is the proton number of each of the incident nuclei and $\tilde{T}%
_{nucl}(\mbox{\boldmath{$\rho$}}^{\prime })$ is the renormalized thickness
function defined in eq.(\ref{renor-thick}). For the penetrating systems, we
have 
\begin{equation}
\phi _{pen}(\tau ,\mbox{\boldmath{$\rho$}},\eta )\!\!=e\,\langle Z^{\prime
}\rangle \!\!\int \!\!d\mbox{\boldmath{$\rho$}}^{\prime }~\tilde{T}_{pen}(%
\mbox{\boldmath{$\rho$}}^{\prime })\sum_{\eta ^{\prime }=\pm ~y_{be}}\left[ ~%
\frac{\cosh \eta ^{\prime }\times \theta (\tau -|\mbox{\boldmath{$\rho$}}-%
\mbox{\boldmath{$\rho$}}^{\prime }|)}{\sqrt{\tau ^2\sinh ^2(\eta -\eta
^{\prime })+|\mbox{\boldmath{$\rho$}}-\mbox{\boldmath{$\rho$}}^{\prime }|^2}}%
~\right] \;  \label{scl-pene}
\end{equation}
and 
\begin{equation}
A_{pen}(\tau ,\mbox{\boldmath{$\rho$}},\eta )\!\!=e\!\!\ \langle Z^{\prime
}\rangle \!\!\int \!\!d\mbox{\boldmath{$\rho$}}^{\prime }~\tilde{T}_{pen}(%
\mbox{\boldmath{$\rho$}}^{\prime })\!\sum_{\eta ^{\prime }=\pm
~y_{be}}\left[ ~\frac{\sinh \eta ^{\prime }\times \theta (\tau -|%
\mbox{\boldmath{$\rho$}}-\mbox{\boldmath{$\rho$}}^{\prime }|)}{\sqrt{\tau
^2\sinh ^2(\eta -\eta ^{\prime })+|\mbox{\boldmath{$\rho$}}-%
\mbox{\boldmath{$\rho$}}^{\prime }|^2}}~\right] ~\!\!,  \label{vec-pene}
\end{equation}
where $\langle Z^{\prime }\rangle $ is the average number of protons which
are included in each penetrating system and $\tilde{T}_{pen}(%
\mbox{\boldmath{$\rho$}}^{\prime })$ is the renormalized thickness function
defined in eq.(\ref{renor-thick}).

\subsection{Total electromagnetic field in high-energy heavy-ion collisions}

The total scalar and vector potentials produced in high-energy heavy-ion
collisions are now written as the following sums: 
\begin{eqnarray}
&&\phi (\tau ,\mbox{\boldmath{$\rho$}},\eta )=\phi _{fluid}(\tau ,%
\mbox{\boldmath{$\rho$}},\eta )+\phi _{nucl}(\tau ,\mbox{\boldmath{$\rho$}}%
,\eta )+\phi _{pen}(\tau ,\mbox{\boldmath{$\rho$}},\eta ), \\
&&A_z(\tau ,\mbox{\boldmath{$\rho$}},\eta )=A_{fluid}(\tau ,%
\mbox{\boldmath{$\rho$}},\eta )+A_{nucl}(\tau ,\mbox{\boldmath{$\rho$}},\eta
)+A_{pen}(\tau ,\mbox{\boldmath{$\rho$}},\eta ).
\end{eqnarray}
Then, it is straightforward to obtain the corresponding electromagnetic
fields by 
\begin{eqnarray}
&&\mbox{\boldmath{$E$}}(\mbox{\boldmath{$x$}},t)=-\mbox{\boldmath{$\nabla$}}%
\phi (\mbox{\boldmath{$x$}},t)-\frac \partial {\partial t}%
\mbox{\boldmath{$A$}}(\mbox{\boldmath{$x$}},t)~, \\
&&\mbox{\boldmath{$B$}}(\mbox{\boldmath{$x$}},t)=\mbox{\boldmath{$\nabla$}}%
\times \mbox{\boldmath{$A$}}(\mbox{\boldmath{$x$}},t)~
\end{eqnarray}
and they are explicitly given by 
\begin{eqnarray}
&&E_x(\tau ,\mbox{\boldmath{$\rho$}},\eta )=\int \!\!d\mbox{\boldmath{$%
\rho$}}^{\prime }\int \!\!d\eta ^{\prime }~\frac{{\cal F}(\eta ^{\prime },%
\mbox{\boldmath{$\rho$}}^{\prime })~\cosh \eta ^{\prime }}{[~\tau ^2\sinh
^2(\eta -\eta ^{\prime })+|\mbox{\boldmath{$\rho$}}-\mbox{\boldmath{$\rho$}}%
^{\prime }|^2~]^{3/2}}\times (\mbox{\boldmath{$\rho$}}-\mbox{\boldmath{$%
\rho$}}^{\prime })_x \\
&&E_y(\tau ,\mbox{\boldmath{$\rho$}},\eta )=\int \!\!d\mbox{\boldmath{$%
\rho$}}^{\prime }\int \!\!d\eta ^{\prime }~\frac{{\cal F}(\eta ^{\prime },%
\mbox{\boldmath{$\rho$}}^{\prime })~\cosh \eta ^{\prime }}{[~\tau ^2\sinh
^2(\eta -\eta ^{\prime })+|\mbox{\boldmath{$\rho$}}-\mbox{\boldmath{$\rho$}}%
^{\prime }|^2~]^{3/2}}\times (\mbox{\boldmath{$\rho$}}-\mbox{\boldmath{$%
\rho$}}^{\prime })_y \\
&&E_z(\tau ,\mbox{\boldmath{$\rho$}},\eta )=\int \!\!d\mbox{\boldmath{$%
\rho$}}^{\prime }\int \!\!d\eta ^{\prime }~\frac{{\cal F}(\eta ^{\prime },%
\mbox{\boldmath{$\rho$}}^{\prime })~\sinh (\eta -\eta ^{\prime })}{[~\tau
^2\sinh ^2(\eta -\eta ^{\prime })+|\mbox{\boldmath{$\rho$}}-%
\mbox{\boldmath{$\rho$}}^{\prime }|^2~]^{3/2}}~\times \tau \\
&&B_x(\tau ,\mbox{\boldmath{$\rho$}},\eta )=\int \!\!d\mbox{\boldmath{$%
\rho$}}^{\prime }\int \!\!d\eta ^{\prime }~\frac{-{\cal F}(\eta ^{\prime },%
\mbox{\boldmath{$\rho$}}^{\prime })~\sinh \eta ^{\prime }}{[~\tau ^2\sinh
^2(\eta -\eta ^{\prime })+|\mbox{\boldmath{$\rho$}}-\mbox{\boldmath{$\rho$}}%
^{\prime }|^2~]^{3/2}}\times (\mbox{\boldmath{$\rho$}}-\mbox{\boldmath{$%
\rho$}}^{\prime })_y \\
&&B_y(\tau ,\mbox{\boldmath{$\rho$}},\eta )=\int \!\!d\mbox{\boldmath{$%
\rho$}}^{\prime }\int \!\!d\eta ^{\prime }~\frac{{\cal F}(\eta ^{\prime },%
\mbox{\boldmath{$\rho$}}^{\prime })~\sinh \eta ^{\prime }}{[~\tau ^2\sinh
^2(\eta -\eta ^{\prime })+|\mbox{\boldmath{$\rho$}}-\mbox{\boldmath{$\rho$}}%
^{\prime }|^2~]^{3/2}}\times (\mbox{\boldmath{$\rho$}}-\mbox{\boldmath{$%
\rho$}}^{\prime })_x \\
&&B_z(\tau ,\mbox{\boldmath{$\rho$}},\eta )=0~,
\end{eqnarray}
where 
\begin{eqnarray}
{\cal F}(\eta ^{\prime },\mbox{\boldmath{$\rho$}}^{\prime })\!\! &=&\!e\,%
\bigg[\tilde{T}_{fluid}(\mbox{\boldmath{$\rho$}}^{\prime })\frac{dN_{net}}{%
d\eta ^{\prime }}+\langle Z^{\prime }\rangle \tilde{T}_{pen}(%
\mbox{\boldmath{$\rho$}}^{\prime })\{\delta (\eta ^{\prime }-y_{be})+\delta
(\eta ^{\prime }+y_{be})\}\bigg] \theta (\tau -|\mbox{\boldmath{$\rho$}}-%
\mbox{\boldmath{$\rho$}}^{\prime }|)  \nonumber \\
&&\hspace*{-.65cm}+\ \;e\,\bigg[Z\tilde{T}_{nucl}(\mbox{\boldmath{$\rho$}}%
^{\prime })\{\delta (\eta ^{\prime }-y_{be})+\delta (\eta ^{\prime
}+y_{be})\}\bigg] \theta (|\mbox{\boldmath{$\rho$}}-\mbox{\boldmath{$\rho$}}%
^{\prime }|-\tau )~~.
\end{eqnarray}

\subsection{Space-time structure of the electromagnetic field}

In order to visualize the space-time distribution of the electromagnetic
field in high-energy heavy-ion collisions, we show now some results of
numerical calculations by means of eqs.(36)-(41). First, Fig.8 shows the
transverse component of the electric field, $E_\rho \;(\equiv \sqrt{%
E_x^2+E_y^2})$, as function of $\rho \;(\equiv \sqrt{x^2+y^2})$ at $t=15$ fm
in the center of mass system. One can see a high peak in the electric field
which increases as $z\rightarrow t$, $i.\,e.$, as the proper time $\tau
\rightarrow 0$. This peak is produced by one of the incident nuclei before
the collision. For smaller values of $z$, the peak which appears close to $%
\rho \sim R_{Pb}$ (radius of Pb) is produced by the expanding charged fluid.
The time dependence of $E_\rho $ at $\rho =5$ fm for various $z$ are shown
in Fig.9. One can observe the propagation of a sharp peak produced by one of
the incident nuclei. The apparent discrepancy of the result for $z=0$ is due
to the fact that both nuclei contribute to the field in this case. It is
also found that the width of the peak is about $\Delta\tau\sim3\,$fm.

The longitudinal component of the electric field $E_z$ is shown in Fig.10.
The magnitude of $E_z$ is much smaller than that of $E_\rho$. At $z$=0, $%
E_z\equiv $0 because of the symmetry. Except for the case of $z=14\,$fm at
small $\rho$, the sign of $E_z$ is negative mainly due to the charge
concentration at large-$\eta$ region (see Fig. 7). The positive $E_z$ in $%
z=14\,$fm is due to one of the incident nuclei.

The magnetic field at $t=15\,$fm is shown in Fig. 11. Because of the
expansion of a positively charged matter, the sign of $B_{\varphi}$ at $z>0$
is positive. One can also roughly estimate the maximum value of the
electromagnetic energy density $\varepsilon_{max}$. The maximum value of $%
E_{\rho}\approx$ 7.5 [MeV/fm] is reached on $z\sim$ 0 fm and $\rho\sim
R_{Pb} $ at $t\sim1\,$fm. Then we have $\varepsilon_{max} \approx$ 0.3 MeV/fm%
$^3$.

\section{Freeze-out surface}

To obtain the freeze-out surface, we use the particle density in the fluid
parametrized as described in Section I$\!$I$\!$I, 
\begin{eqnarray}
n(\tau ,\eta ,\rho )=\tilde{T}_{fluid}(\rho )~\frac{dN_{fluid}}{d\eta }%
\times \frac 1\tau ~,  \label{parden}
\end{eqnarray}
where $dN_{fluid}/d\eta $ is constant in some rapidity interval. Since the
most of the produced particles are pions, let us reinterpret it as the
portion which corresponds to pions and fix the parameters by fitting the
data on

\begin{eqnarray}
\frac{dN_\pi }{dy}\approx ~3\times \!\frac{dN_{\pi ^{-}}}{dy},
\label{real-pi}
\end{eqnarray}
where $dN_\pi /dy$ is related with $dN_{fluid}/d\eta $ through 
\begin{eqnarray}
\frac{dN_\pi }{dy}\simeq {\cal N}\int \!\!d\eta \frac{dN_{fluid}}{d\eta }%
\times \frac 1{e^{\langle m_{\mbox{\tiny T}}\rangle \cosh (y-\eta )/T}-1}.
\label{eq:fluid}
\end{eqnarray}
Here we are including only pions, considering the other hadrons as
contaminations. The NA49's data for negative hadrons include $\bar{p}$ and $%
K^{-}$ \cite{NA49-Appelshauser}. Using the experimental results in the mid
rapidity region, $\langle K^{+}+K^{-}+2K_s^0\rangle /\langle \pi ^{+}+\pi
^0+\pi ^{-}\rangle \approx $ 0.135, $K^{+}/K^{-}\approx $1.8 and $%
K_s^0/K^{-}\approx $1.8 \cite{NA49-Bormann}, one can estimate the
multiplicity of kaons as shown in Table. \ref{table-centrality} and the
total pion multiplicity (including $\pi ^0$) is 1950 and 1640 for the
centrality 5\% and 15\%, respectively. The solid line in Fig.12 shows the
total pion rapidity distribution for centrality 15\%. The shaded area and
the dashed curve are our parametrization for $dN_{fluid}/d\eta $ and $dN_\pi
/dy$ respectively.

Now, we can compute the freeze out surface by equating the density given in
eq.(\ref{parden}) to the critical value $n_f$ for the Bose-Einstein
distribution at the freeze-out temperature $T_f$ 
\begin{eqnarray}
\tilde{T}_{fluid}(\rho )\frac{dN_{fluid}}{d\eta }\bigg|_{\eta =0}\frac 1\tau
=n_f\equiv \frac{3m_\pi ^2~T_f}{2\pi ^2}\sum_{n=0}^\infty \frac{K_2[(n+1){%
\frac{m_\pi }{T_f}}]}{n+1}~.  \label{surface}
\end{eqnarray}
For $T_f$=140 MeV, one obtains $n_f$= 0.096 fm$^{-3}$. The intercept of the
freeze-out surface in ($\tau ,\rho $)-plane is shown in Fig.13.

\section{Single-particle spectra and Coulomb effect\label{Six}}

\subsection{Transverse-momentum spectra of charged particles}

Now that the electromagnetic field and the freeze-out surface have been
obtained, one can solve the equation of motion for a classical charged
particle, 
\begin{eqnarray}
\frac d{dt}\frac{m\mbox{\boldmath{$v$}}(t)}{\sqrt{1-\mbox{\boldmath{$v$}}%
^2(t)}}=q\left[ \mbox{\boldmath{$E$}}(\mbox{\boldmath{$x$}},t)+%
\mbox{\boldmath{$v$}}(t)\!\times \mbox{\boldmath{$B$}}(\mbox{\boldmath{$x$}}%
,t)\right] ~,  \label{eq-motion}
\end{eqnarray}
starting from the freeze-out surface. In the present work, we have neglected
the effect of the Coulomb field before the freeze out. The time evolution of
the $\pi ^{+}$- and the $\pi ^{-}$- momenta, as given by eq.(\ref{eq-motion}%
), can generally be expressed in terms of the location and the initial
momentum $m_{\mbox{\tiny T}0}$ at the freeze out as 
\begin{eqnarray}
m_{\mbox{\tiny T}}^{\pm }=m_{\mbox{\tiny T}}^{\pm }(\rho ,\eta ;m_{%
\mbox{\tiny T}0},y_0,\varphi ),  \label{mt-shift}
\end{eqnarray}
where {\bf $\rho ,\eta ,m_{\mbox{\tiny T}0},y_0,\varphi $ }are the phase
space coordinates of {\bf $\pi $ }at the freeze-out time{\bf . }We show in
Fig.14 some results for the specified initial conditions. One can see that $%
\pi ^{+}$s are accelerated outward and $\pi ^{-}$s are decelerated by the
electromagnetic field. The rapidity {\bf $y_0$} also shifts to {\bf $y_{\pm
} $ .} However, this change is quite small, {\bf $|y_{\pm }-y_0|\ll $ }%
0.1\thinspace .

Since the particle number density is given by eq.(\ref{parden}), if we
neglect the Coulomb effect, the invariant cross section for $\pi ^{\pm ,0}$
would be written in the form{\bf \ 
\begin{eqnarray}
E\,\frac 1{\sigma _{tot}}\frac{d^3\sigma }{d^3\mbox{\boldmath{$p$}}}\!\!\;
&=&\;\!\!\frac 1{(2\pi )^3}\int \hspace*{-2mm}\int \hspace*{-2mm}\int %
\hspace*{-2mm}\int_{F.O.surface}\hspace*{-4mm}p_\mu d\sigma ^\mu %
\hspace*{+1mm}\frac{1}{e^{m_{\mbox \tiny T0}\cosh (\eta -y)/T_f}-1}
  \nonumber \\
\!\! &=&\;\!\!\frac 1{(2\pi )^3}
\int\!\!d\phi\int_{\eta _{min}}^{\eta _{max}}
\hspace*{-5mm}d\eta~ \left\{~
\int \!\!\rho d\rho ~\tau _f(\rho )~m_{\mbox{\tiny T0}}
\cosh (\eta-y) 
+ \int\!\! d\tau \tau\rho_f(\tau) ~p_{\mbox{\tiny T0}}
  \cos(\varphi-\phi) ~\right\} 
~\frac{1}{e^{m_{\mbox \tiny T0}\cosh (\eta -y)/T_f}-1}~.
\label{inv-cross}
\end{eqnarray}
}So, the transverse momentum distribution in this case is {\bf 
\begin{eqnarray}
\frac{dN}{m_{\mbox{\tiny T}0}dm_{\mbox{\tiny T}0}} &=& 
\frac 1{(2\pi )^2}%
\int\hspace*{-2.2mm}\int dyd\varphi \int_{\eta _{min}}^{\eta _{max}}~
\hspace*{-5mm}d\eta ~\left\{~
\int \!\!\rho d\rho ~\tau _f(\rho )~m_{\mbox{\tiny T}0}
  \cosh (\eta -y)
  +\int\!\! d\tau \tau\rho_f(\tau) ~p_{\mbox{\tiny T0}}
  \cos\varphi ~\right\} 
~\frac{1}{e^{m_{\mbox \tiny T0}\cosh (\eta -y)/T_f}-1}~. 
\label{single-spectrum}
\end{eqnarray}
}

Since the transverse mass $m_{\mbox{\tiny T}0}$ at freeze-out time shifts to 
$m_{\mbox{\tiny T}}^{\pm }$ as given by eq.(\ref{mt-shift}), the single
particle spectrum eq.(\ref{single-spectrum}) is distorted for $\pi ^{+}$ as 
\begin{eqnarray}
\frac{dN^{+}}{m_{\mbox{\tiny T}}^{+}dm_{\mbox{\tiny T}}^{+}}\!\!\; &=&\!\!\;%
\frac 1{(2\pi )^2}\int \!\!\!\!\int \!\!dyd\varphi \int_{\eta _{min}}^{\eta
_{max}}\hspace*{-5mm}d\eta ~\int \!\!\rho d\rho ~\tau _f(\rho )\nonumber \\ 
&& \times \int \!\!dm_{\mbox{\tiny T0}}~
   \delta [m_{\mbox{\tiny T0}}-m_{\mbox{\tiny T}}^{+}
   (\rho ,\eta ,m_{\mbox{\tiny T}},y,\varphi )]
    \left\{~ 
    m_{\mbox{\tiny T0}}\cosh (\eta -y)~
  + p_{\mbox{\tiny T0}} \bigg|\frac{\del \tau_f(\rho)}{\del \rho}
    \bigg|\cos\varphi~\right\}~
    \frac{1}{e^{m_{\mbox{\tiny T0}}\cosh (\eta -y)/T_f}-1} 
\nonumber \\
\!\! &=&\!\!\;
\frac 1{(2\pi )^2}\int \!\!\!\!\int \!\!dyd\varphi 
\int_{\eta_{min}}^{\eta _{max}}\hspace*{-5mm}d\eta 
~\int \!\!\rho d\rho ~\tau _f(\rho)
\left| \frac{\partial m_{\mbox{\tiny T0}}}{\partial m_{\mbox{\tiny T}}^{+}}%
\right| _{m_{\mbox{\tiny T0}}=m_{\mbox{\tiny T}}^{+}}
 \hspace*{-12mm} 
 \{~m^+_{\mbox{\tiny T}}\cosh (\eta -y)~
  + p^+_{\mbox{\tiny T}} \bigg|\frac{\del \tau_f(\rho)}{\del \rho}
    \bigg|\cos\varphi~\}~
    \frac{1}{e^{m^+_{\mbox{\tiny T}}\cosh (\eta -y)/T_f}-1} ~. 
\label{Coulomb-distorted}
\end{eqnarray}
For $\pi ^{-}$, one can obtain a similar expression.

However, these are not yet the final results for the charged-particle
spectra. In writing eq.(\ref{Coulomb-distorted}), we have neglected the
possibility of frozen-out particles be reabsorbed by the fluid again. So, we
have checked all the trajectories of $\pi^{\pm}$ mesons after leaving the
freeze-out surface and excluded those ones which crossed it in, considering
them as re-absorbed by the fluid.

\subsection{Yield ratio $\pi^-/\pi^+$}

Using the final results of eq.(\ref{Coulomb-distorted}) which we have just
derived, we obtain a formula for the pionic yield ratio 
\begin{eqnarray}
\pi ^{-}/\pi ^{+}=\frac 1{m_{\mbox{\tiny T}}^{-}}\frac{dN^{-}}{dm_{%
\mbox{\tiny
T}}^{-}}~\bigg/~\frac 1{m_{\mbox{\tiny T}}^{+}}\frac{dN^{+}}{dm_{%
\mbox{\tiny
T}}^{+}}~~\bigg|_{m_{\mbox{\tiny T}}^{-}=m_{\mbox{\tiny T}}^{+}~\equiv m_{%
\mbox{\tiny T}}}~~.
\end{eqnarray}
To compare more precisely our results with the NA44 data \cite{NA44-Coulomb}
for centrality 15\% , we introduce the acceptance correction factor ${\cal A}%
(m_{\mbox{\tiny T}},y)$ to our formula eq.(\ref{Coulomb-distorted}), which
would approximately reproduce the nature of the NA44's detectors: 
\begin{eqnarray}
{\cal A}(m_{\mbox{\tiny T}};y)=\delta (y+1.67\sqrt{m_{\mbox{\tiny T}%
}^2-m_\pi ^2}-4.2)~.
\end{eqnarray}
Here we neglect the rapidity shift due to the Coulomb field.

The results of our calculation together with NA44's data are shown in
Fig.15. As seen, our results approximately reproduce the data. We shall now
look more carefully into our results in Fig.15. It is found that our model
overestimate in the region $m_{\mbox{\tiny T}}$-$m_\pi \sim $ 5MeV and
underestimate in the region 10 $\le m_{\mbox{\tiny T}}$-$m_\pi \le $ 80~MeV.
The underestimation of our model may be improved by introducing a transverse
expansion for the charged fluid, because it makes the freeze-out time
shorter and then the frozen out particles would feel a stronger
electromagnetic field. Another factor which we have neglected in the present
calculation, namely the effect of the field before the freeze out, also
contributes to correct the results in the right direction. It has also been
suggested by other authors\cite{Arbex} that $\pi ^{-}/\pi ^{+}$ ratio has
origin in hyperon decays \footnote{%
Although most of $\pi ^{-}$ from $\Lambda $-decay should appear as small $p_{%
\mbox{\tiny T}}$ particles, in the NA44 Collaboration's setup almost all the
hyperons (with rapidity in the range $y_{lab.}\le $ 2$\sim $3) decay in the
first dipole magnet region\cite{Kaneta}. This means that very few daughter $%
\pi ^{-}$ can enter their detector. So, although we think that this effect
should be studied, it is seemingly very small.
\par
}. 

Save particle creation and annihilation, no quantum effect has been
considered in the present calculation. Due to the uncertainty relation,
evidently the classical calculation only has meaning if the wave packets are
sufficiently small. This implies that the Coulomb effect for small-$p_{%
\mbox{\tiny T}}$ particles\cite{osada96,osada-tohoku} would be smaller,
because such particles would have a large wave packet for a good definition
of the momentum and would feel a weaker field, which is an average field
over their extension. This deviation would be larger for $\pi^-$ than for $%
\pi^+$, because the former is attracted toward the symmetry axis whereas the
latter is repelled away.

We also show our result of $\pi^-/\pi^+$ for the centrality 5\% in Fig.16.
In this calculation, the rapidity of the pions is fixed at $y_{\pi}$=0.0 and
2.0 and set ${\cal A}(m_{\mbox{\tiny T}}; y)\equiv$ 1.

\section{Concluding remarks and discussions}

In this paper, we have investigated the structure of the electromagnetic
fields which accompany high-energy heavy-ion collisions and computed the
Coulomb effect for pion spectra caused as a consequence. We have considered
three possible origins of the total electromagnetic field, $i.$ $e.,$
i)~incident nuclei, ii)~longitudinally expanding charged matter and
iii)~penetrating systems which are composed of surviving protons. Glauber
model has been used to estimate the number of participant nucleons and to
determine the transverse profile of the charged system after the collision.
The centrality dependence of the net-charge is also taken into account. In
the description of the space-time evolution of the charged matter, a
parametrization based on one-dimensional hydrodynamical model has been used,
with a constant rapidity $\eta $.

The main results of this paper could be summarized as the following.

\begin{itemize}
\item  Compared with the expanding charged matter, the incident nuclei bring
a rather strong electromagnetic field into the interaction region in $\tau
\sim $ a few fm.

\item  The NA44 Pb+Pb data on $\pi ^{-}/\pi ^{+}$ are approximately
reproduced by our model (especially due to the longitudinally expanding
fluid part). This means that the longitudinally expanding charged matter
plays an important role on the final-state interaction of produced hadrons.

\item  As expected, the penetrating systems seem to be not important for the
observables in the mid-rapidity region. (However, it should be noted that
they could have non-negligible contributions in the fragmentation regions.)
\end{itemize}

In addition to these points, we would like to note that the Coulomb effect
seen in the yield ratio $\pi ^{-}/\pi ^{+}$ may depend both on the dumping
speed of the electromagnetic field and on the cooling speed of the hadronic
matter. This point leads us further into consideration of the Coulomb effect
which could be observed in the RHIC energies. One would expect that the
strength of Coulomb effect may be small, because the fluid takes much time
to cool down before the freeze out begins and the net-charge density in mid
rapidity region become small at that time. To see this quantitatively, we
present the prediction for the Coulomb effect in Au+Au collisions
(centrality 5\%) at 100GeV+100GeV energy in Fig.19. The net-charge
distribution and the fluid distribution used in this calculation are shown
in Fig.17 and Fig.18, respectively. For the net-charge distribution, we
assume that the total net charge and width of `shoulders' at large $|\eta |$
regions is to be the same as that of SPS case\footnote{%
We also calculated $\pi ^{-}/\pi ^{+}$ by using a somewhat different net
charge distribution, $dN_{fluid}/d\eta\sim$ 10, for the central region.
However, no considerable difference was obtained.}.The total number of pions
in the RHIC energy in Fig.18 is determined by the energy conservation law
for the total system. As shown in Fig.19, the enhancement of $\pi ^{-}/\pi
^{+}$ is seen at very small $m_{\mbox{\tiny T}}$ -$m_\pi $ region only.

As remarked in Section V$\!$I, there remained some open questions: effect of
the transverse expansion, charge redistribution inside the fluid,
introduction of the resonance decays and the effect of the uncertainty
relation. These questions are presently under investigation.

\bigskip
\bigskip

\noindent{\bf Acknowledgments}

This work has been partially supported by FAPESP under the contracts
98/2463-2 and 98/00317-2. The authors acknowledge stimulating discussions
with T. Kodama, H-T. Elze, F. Grassi and S. Padula. We are especially
indebted to G. Odyniec (NA49 Collab.) and M. Kaneta (NA44 Collab.){\bf \ }%
for clarification of data and experimental set up. One of the authors (T.
O.) would like to thank M. Maruyama, F. Takagi and M. Biyajima for helpful
discussions and encouragement and wishes to express gratitude to M. Ueda for
helpful suggestions and hospitable support at S\~{a}o Paulo.

\clearpage 
\begin{table}[tbp]
\begin{center}
\begin{tabular}{|l||ll|ll|}
\hline
\multicolumn{5}{|c|}{Pb+Pb 158\thinspace GeV/nucleon} \\ \hline\hline
& \multicolumn{2}{|c|}{5\thinspace \% centrality ($b\le $3.5 fm)} & 
\multicolumn{2}{c|}{15\thinspace\% centrality ($b\le $6.1 fm)} \\ \hline
$N^{part}$ & 352 & $N_{B-\bar{B}}$ in Ref.\cite{NA49-Appelshauser} & 298 & 
Glauber \\ 
$p-\bar{p}$ & 148.9 & Ref.\cite{NA49-Appelshauser} & 125.5 & Glauber \\ 
$h^{-}\equiv \pi ^{-},K^{-},\bar{p}$ & 695 & Ref.\cite{NA49-Appelshauser} & 
585.9 & Glauber \\ 
$\Lambda -\bar{\Lambda}$ & 27.4 & Ref.\cite{NA49-Bormann} & 22.5 & Ref.\cite
{Andersen98}+Glauber \\ 
$\bar{\Lambda}$ & 7.5 & Ref.\cite{NA49-Bormann} & 5.6 & Ref.\cite{Andersen98}%
+Glauber \\ 
$\bar{p}$ & 4.4 & empirical & 3.5 & Ref.\cite{Andersen98}+Glauber+empirical
\\ \hline\hline
$N_{h^{-}-h^{+}}$ & 10.0 &  & 10.6 &  \\ 
$N_{net}=2Z$ & 139 &  & 116 &  \\ 
$<Z^{\prime }>$ & 12.5 & single hemisphere & 24.6 & single hemisphere \\ 
\hline\hline
$K^{+}$ & 81 & \cite{NA49-Bormann,NA49-Appelshauser} & 68 & Glauber \\ 
$K^{-}$ & 45 & \cite{NA49-Bormann,NA49-Appelshauser} & 38 & Glauber \\ \hline
\end{tabular}
\end{center}
\caption{}
\label{table-centrality}
\end{table}


\noindent 
{\large {\bf Table caption}}

\begin{description}
\item  {\bf Table I:~} Centrality dependence of the participant nucleon
numbers, hadron multiplicities and net charge excess in Pb+Pb collision at
158\thinspace GeV/nucleon.
\end{description}

\clearpage 

\noindent 
{\large {\bf Figure captions}}

\begin{description}
\item  {\bf Fig.1:~} $h$-tube interaction in high-energy $h$-$A$ collisions.

\item  {\bf Fig.2:~} $A$-$B$ collision with impact parameter $%
\mbox{\boldmath{$b$}}$ as seen in the transverse ($x$-$y$) plane.

\item  {\bf Fig.3:~} Transverse profile of the average nucleon density for
the fluid, penetrating systems and incident nuclei, respectively. The
results are shown for Pb+Pb collisions (centrality 5\%).

\item  {\bf Fig.4:~} Net proton ($p$-$\bar{p}$) distribution in Pb+Pb
158\thinspace GeV/nucleon.

\item  {\bf Fig.5:~} Negative hadron ($\pi ^{-}$, $K^{-}$ and $\bar{p}$)
distribution in Pb+Pb 158\thinspace GeV/nucleon.

\item  {\bf Fig.6:~} $\overline{\Lambda }$ distribution and estimated $\bar{p%
}$ distribution in Pb+Pb 158\thinspace GeV/nucleon.

\item  {\bf Fig.7:~} Net-charge distribution of the fluid (shaded area). The
dashed curve is the result of convolution, eq.(19). The solid curve is the
charged particle distribution, eq.(18).

\item  {\bf Fig.8:~} Electric field $E_\rho $ at $t$=15\thinspace fm.

\item  {\bf Fig.9:~} Time dependence of $E_\rho $ at $\rho $=5\thinspace fm.

\item  {\bf Fig.10:~} Longitudinal component of the electric field $E_z$ at $%
t$=15\thinspace fm.

\item  {\bf Fig.11:~} Magnetic field $B_\varphi $ at $t$=15\thinspace fm.

\item  {\bf Fig.12:~} Total pion distribution and fluid distribution $%
dN_{fluid}/d\eta $.

\item  {\bf Fig.13:~} Freeze-out surface projected on the $\tau $-$\rho $
plane. Due to our parametrization, eq.(\ref{fluid-para}), the curve in $\tau 
$-$\rho $ plane is independent of $\eta $.

\item  {\bf Fig.14:~} Time evolution of the momentum of $\pi ^{+}$(solid
line) and $\pi ^{-}$(dashed line). Initial conditions are chosen as $x$%
=6\thinspace fm, $y$=$z$=0\thinspace fm at $t=15\,$fm with momentum, 20, 40,
60 and 80\thinspace MeV in the $+x$ direction(transverse radial direction).

\item  {\bf Fig.15:~} Comparison of our result for the yield ratio $\pi
^{-}/\pi ^{+}$ (solid line) with NA44 Collab's data\cite{NA44-Coulomb}.
Remark that we did not fit the data, but the normalization of the curve has
been done by estimating the overall ratio through data.

\item  {\bf Fig.16:~} Our prediction of $\pi ^{-}/\pi ^{+}$ in Pb+Pb
158GeV/nucleon for the centrality 5\%. The pion rapidity $y_\pi $ is fixed
to be 0.0 and 2.0.

\item  {\bf Fig.17:~} Net charge distribution at a RHIC energy

\item  {\bf Fig.18:~} Fluid distribution at a RHIC energy. The pion
multiplicity is determined by the energy conservation law for the total
system.

\item  {\bf Fig.19:~} Prediction for $\pi ^{-}/\pi ^{+}$ at a RHIC energy.
\end{description}
\clearpage 
\begin{figure}[tbh]
{\vspace*{1.0cm} {\hspace*{+2.0cm} \epsfysize=7.0cm %
\epsfbox{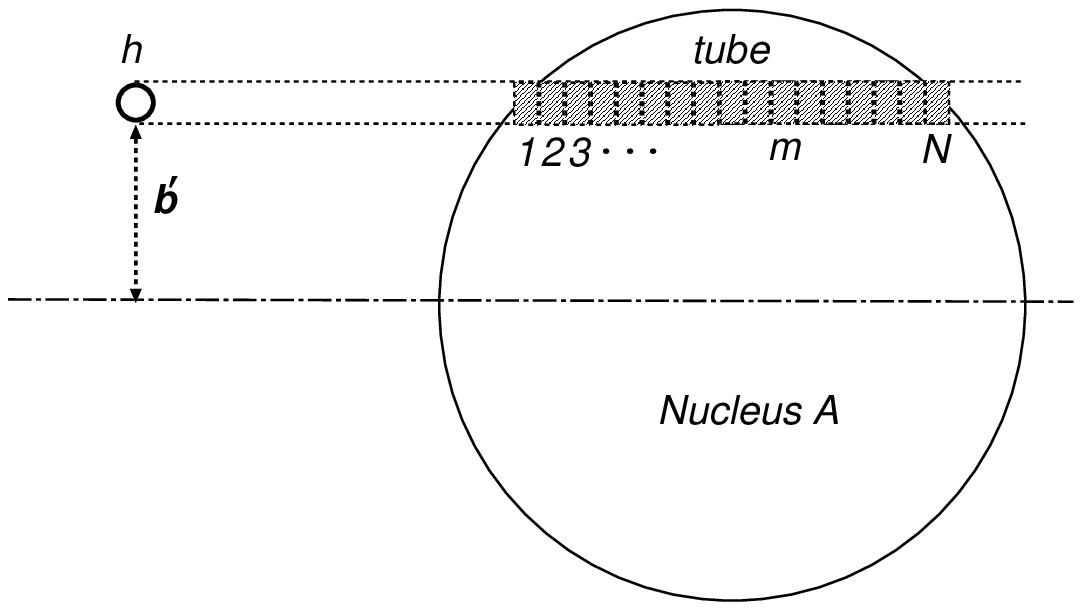}}}
\caption{}
\end{figure}
\begin{figure}[tbh]
{\vspace*{1.0cm} {\hspace*{+2.0cm} \epsfysize=9.0cm %
\epsfbox{Fig2.eps}}}
\caption{}
\end{figure}
\begin{figure}[tbh]
{\vspace*{1.0cm} {\hspace*{+2.0cm} \epsfysize=9.5cm %
\epsfbox{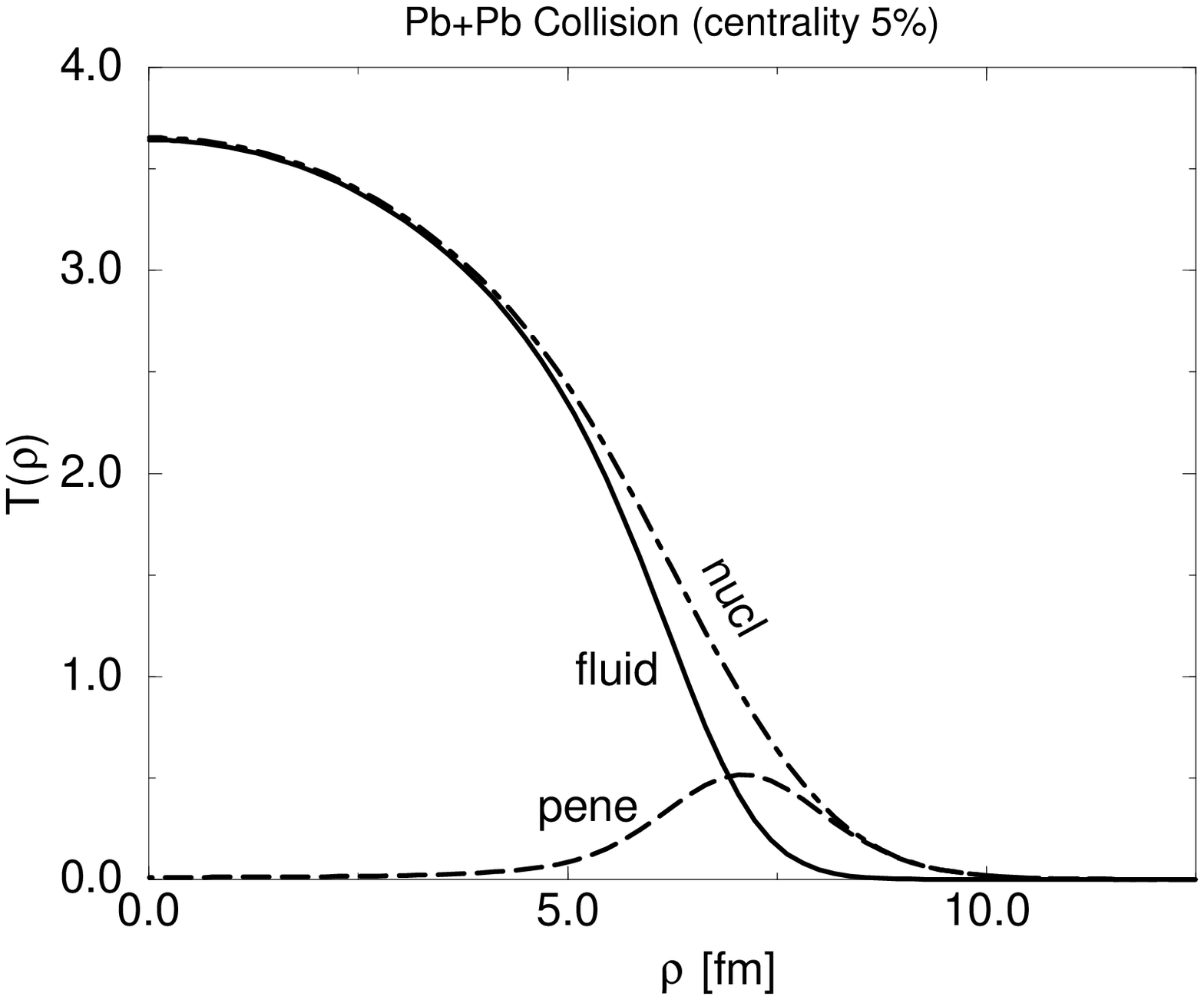}}}
\caption{}
\end{figure}
\begin{figure}[tbh]
{\vspace*{1.0cm} {\hspace*{+2.0cm} \epsfysize=9.5cm %
\epsfbox{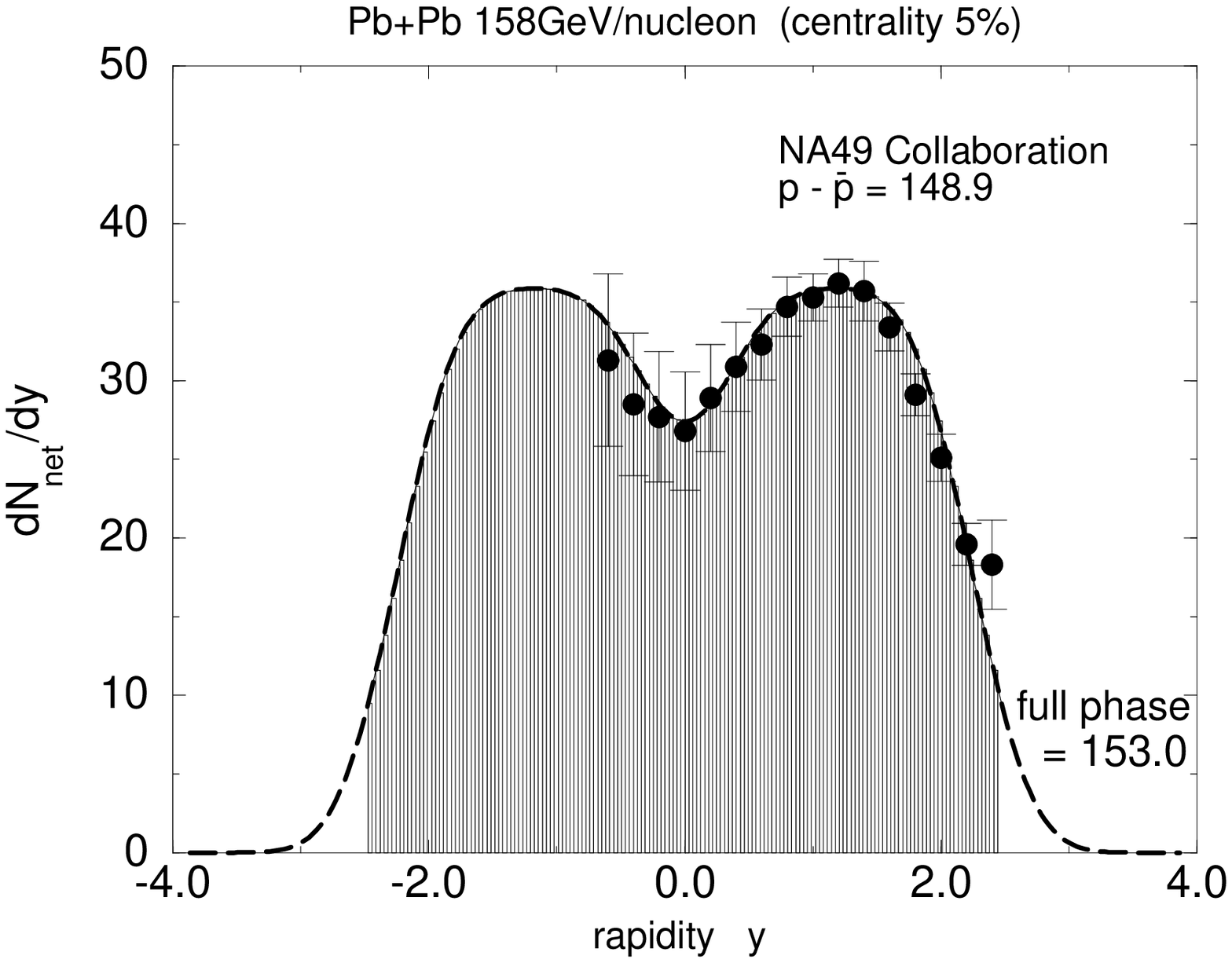}}}
\caption{}
\end{figure}
\begin{figure}[tbh]
{\vspace*{1.0cm} {\hspace*{+2.0cm} \epsfysize=9.5cm %
\epsfbox{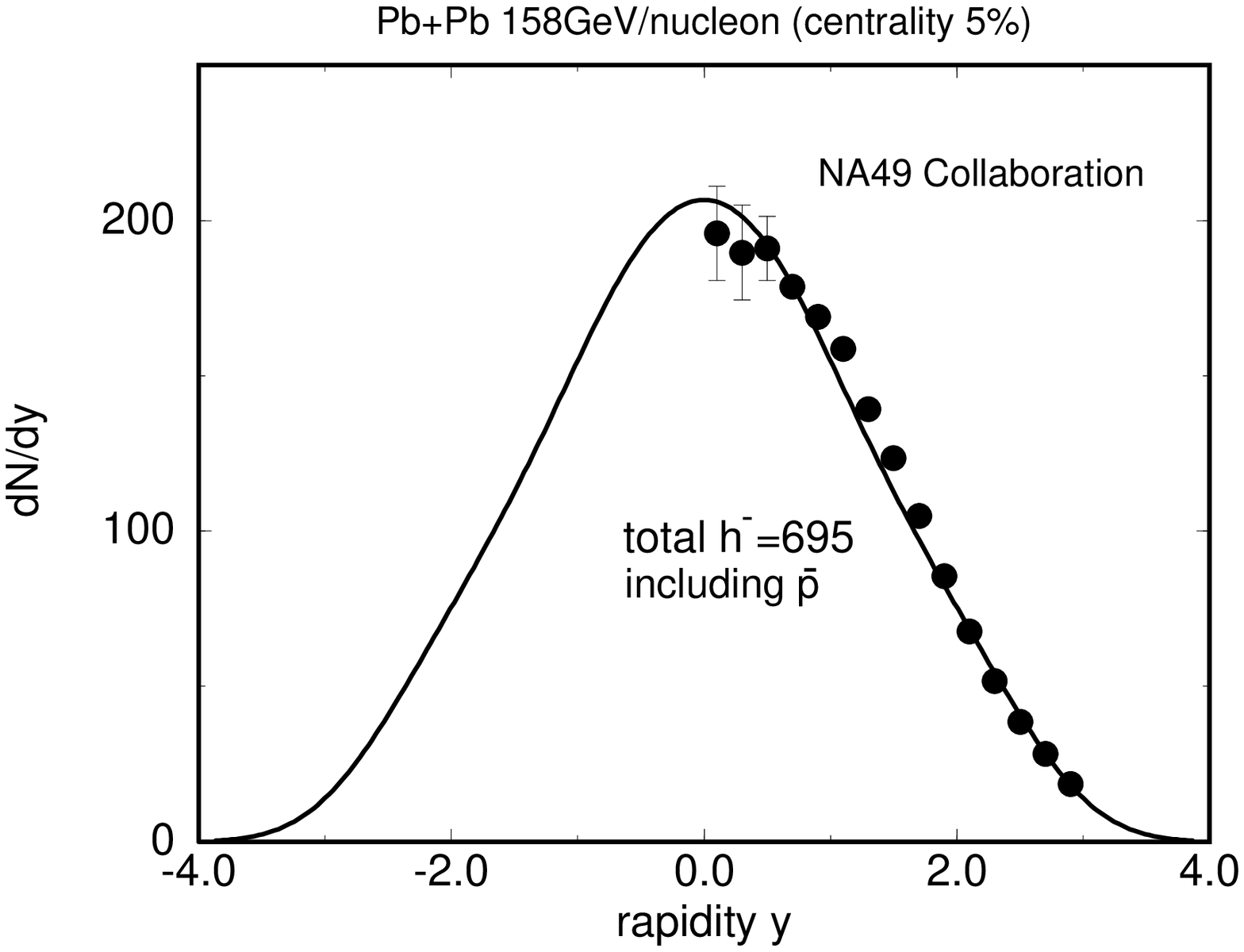}}}
\caption{}
\end{figure}
\begin{figure}[tbh]
{\vspace*{1.0cm} {\hspace*{+2.0cm} \epsfysize=9.5cm %
\epsfbox{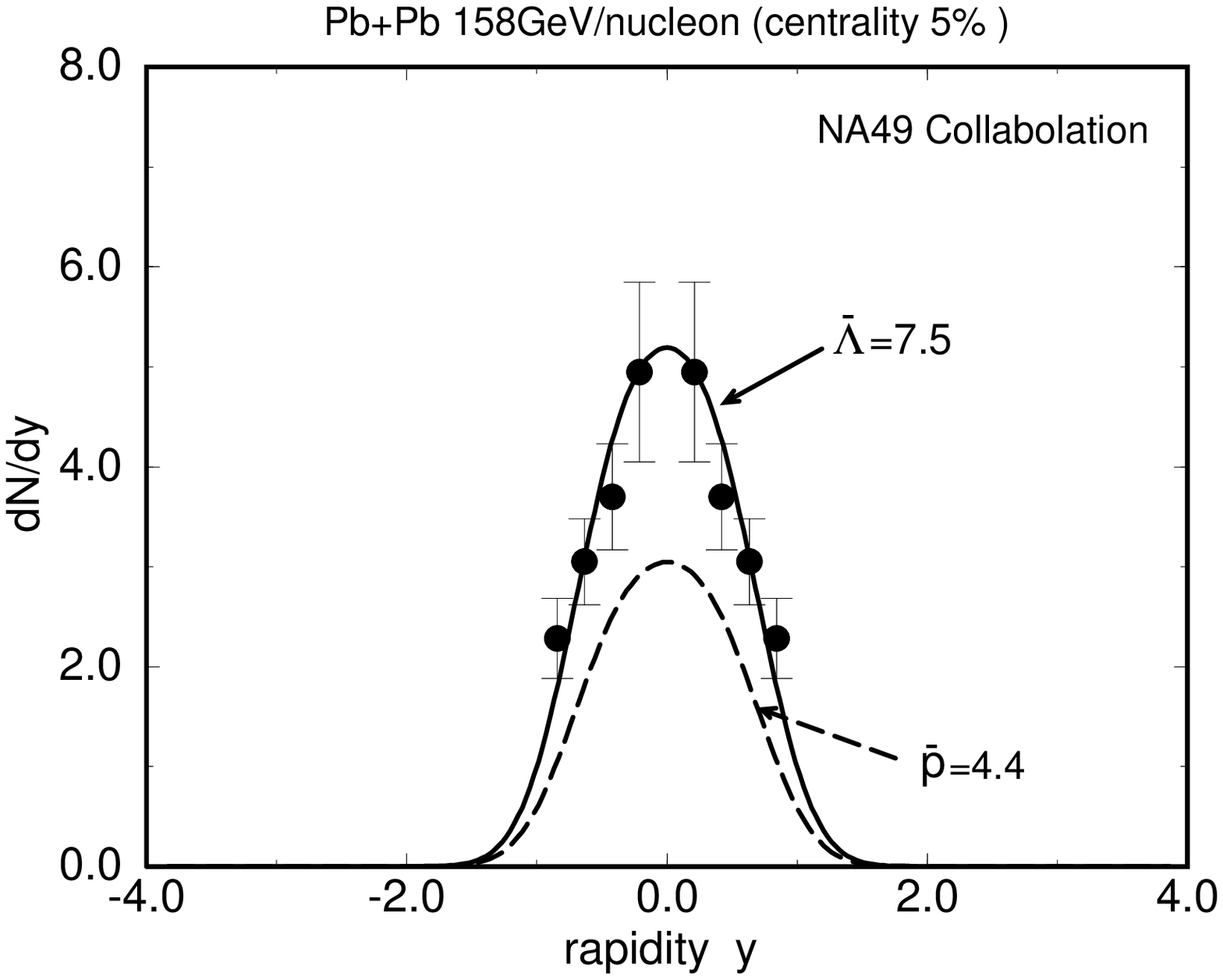}}}
\caption{}
\end{figure}
\begin{figure}[tbh]
{\vspace*{1.0cm} {\hspace*{+2.0cm} \epsfysize=9.5cm %
\epsfbox{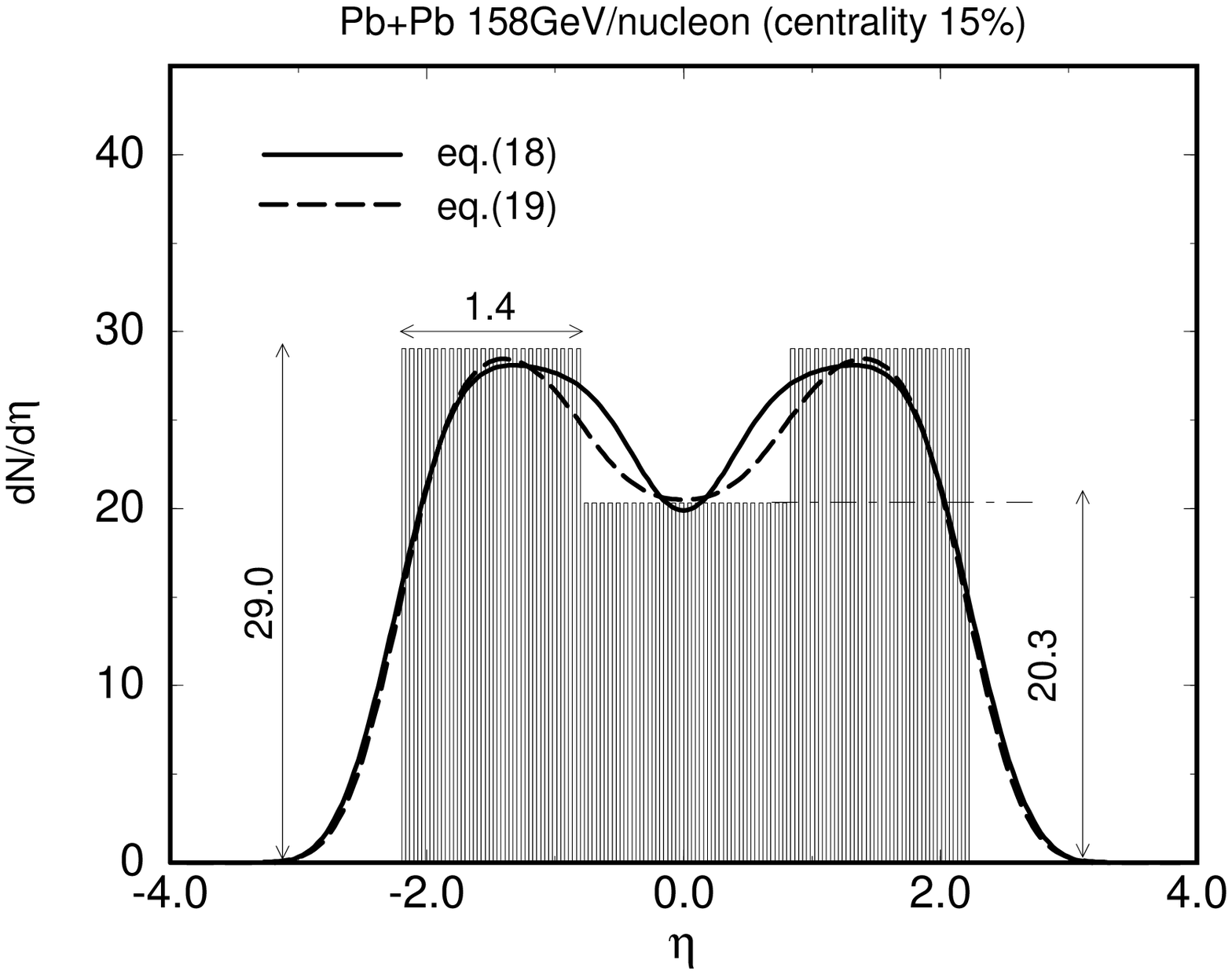}}}
\caption{}
\end{figure}
\begin{figure}[tbh]
{\vspace*{1.0cm} {\hspace*{+2.0cm} \epsfysize=9.5cm %
\epsfbox{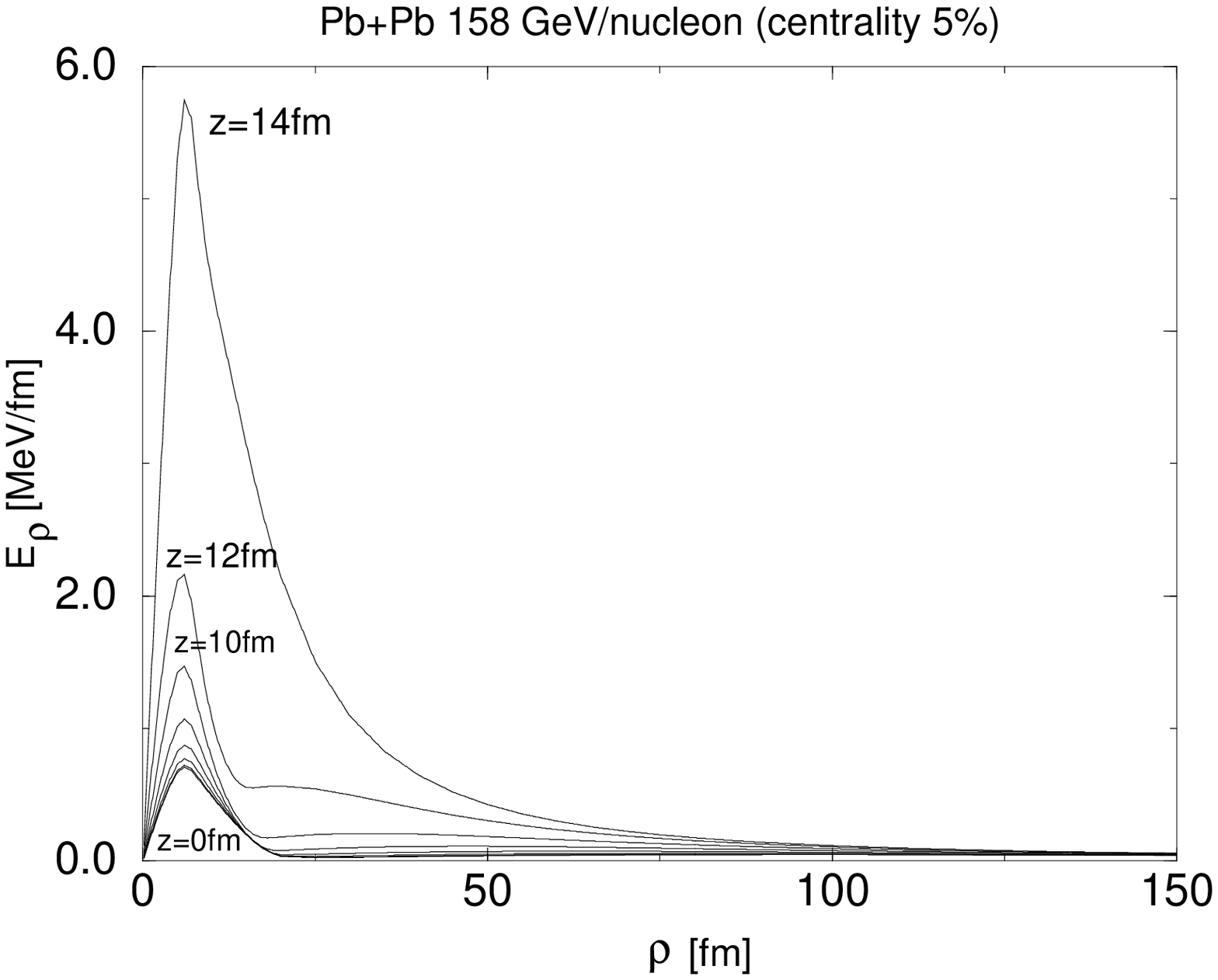}}}
\caption{}
\end{figure}
\begin{figure}[tbh]
{\vspace*{1.0cm} {\hspace*{+2.0cm} \epsfysize=9.5cm %
\epsfbox{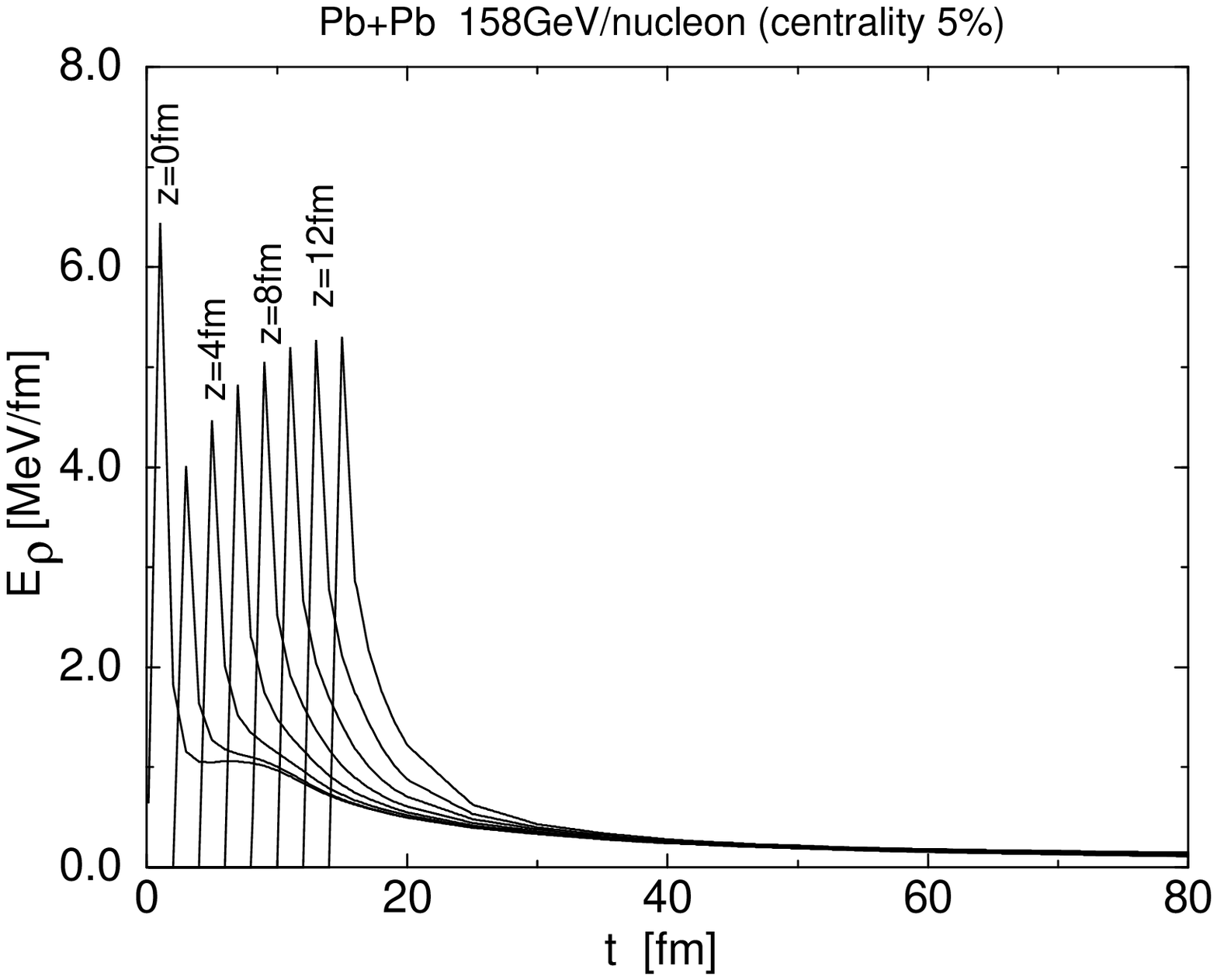}}}
\caption{}
\end{figure}
\begin{figure}[tbh]
{\vspace*{1.0cm} {\hspace*{+2.0cm} \epsfysize=9.5cm %
\epsfbox{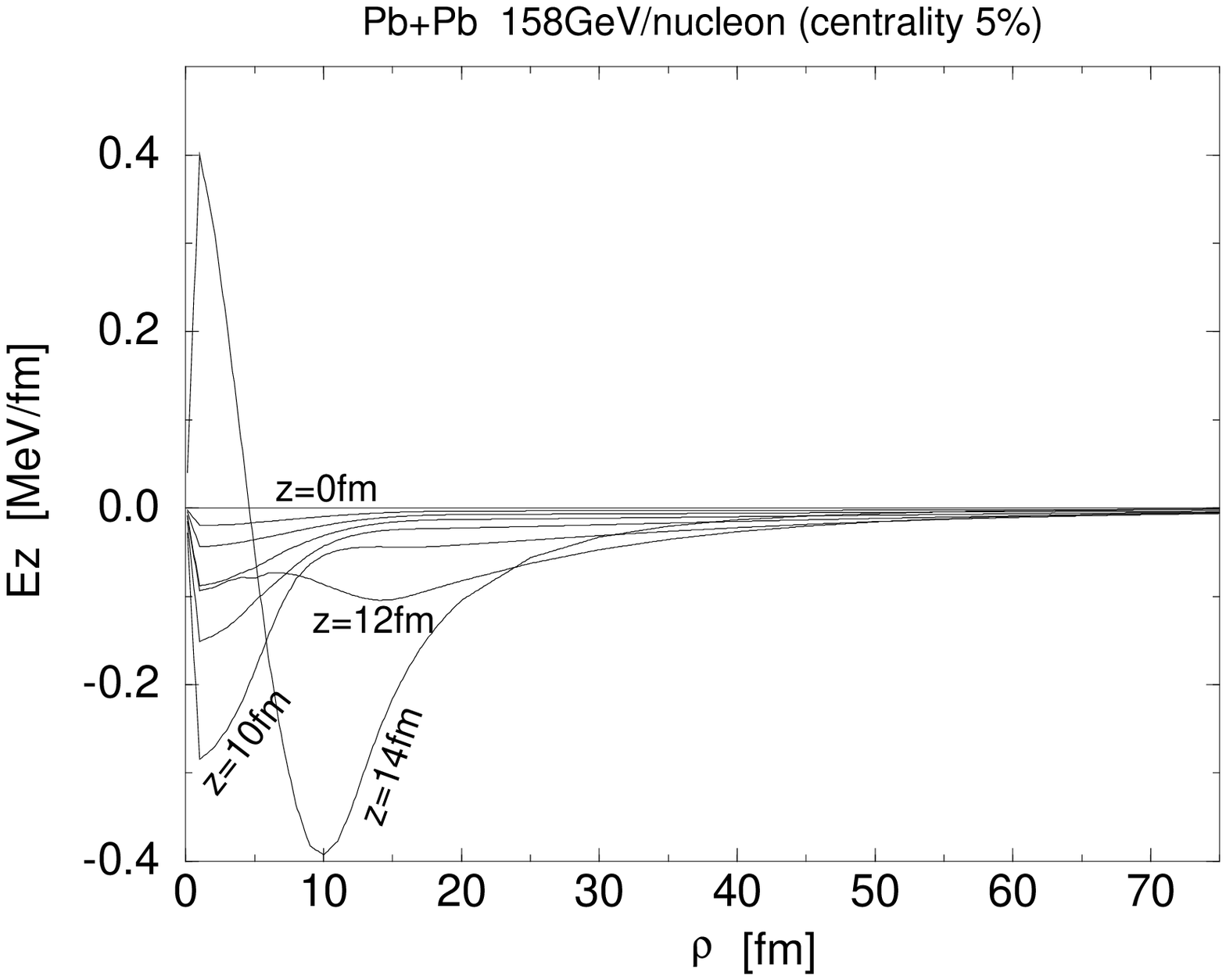}}}
\caption{}
\end{figure}
\begin{figure}[tbh]
{\vspace*{1.0cm} {\hspace*{+2.0cm} \epsfysize=9.5cm %
\epsfbox{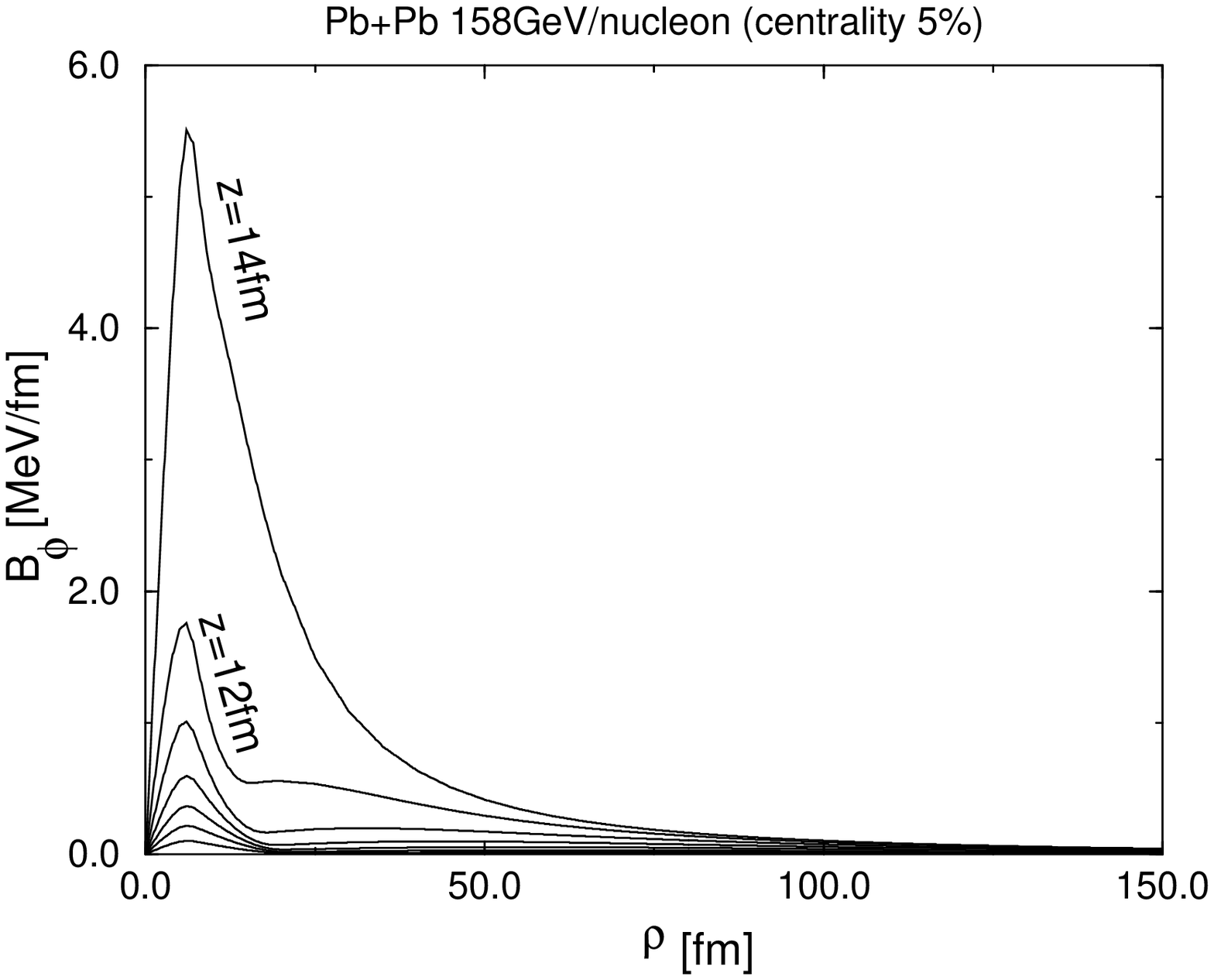}}}
\caption{}
\end{figure}
\begin{figure}[tbh]
{\vspace*{1.0cm} {\hspace*{+2.0cm} \epsfysize=9.5cm %
\epsfbox{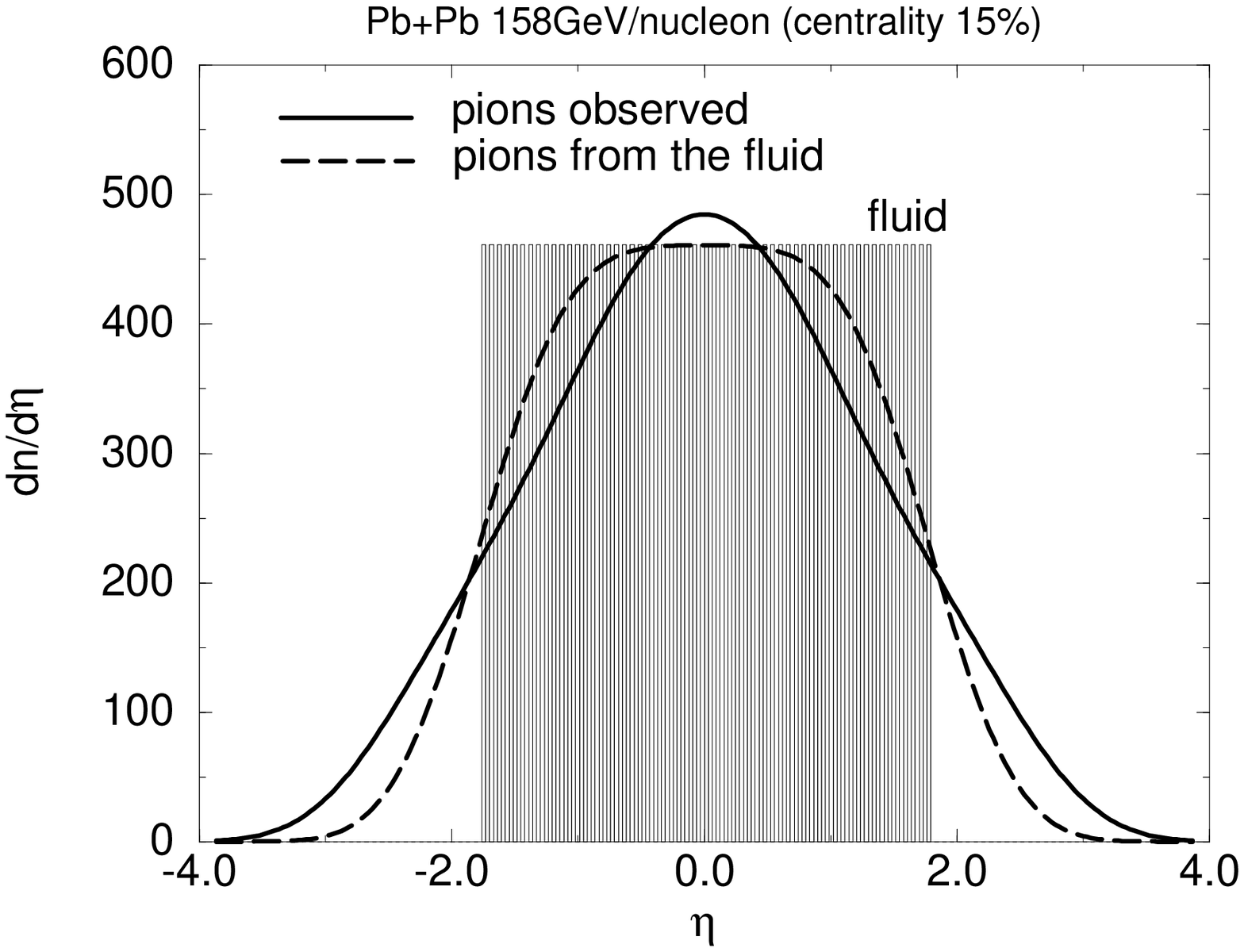}}}
\caption{}
\end{figure}
\begin{figure}[tbh]
{\vspace*{1.0cm} {\hspace*{+2.0cm} \epsfysize=9.5cm %
\epsfbox{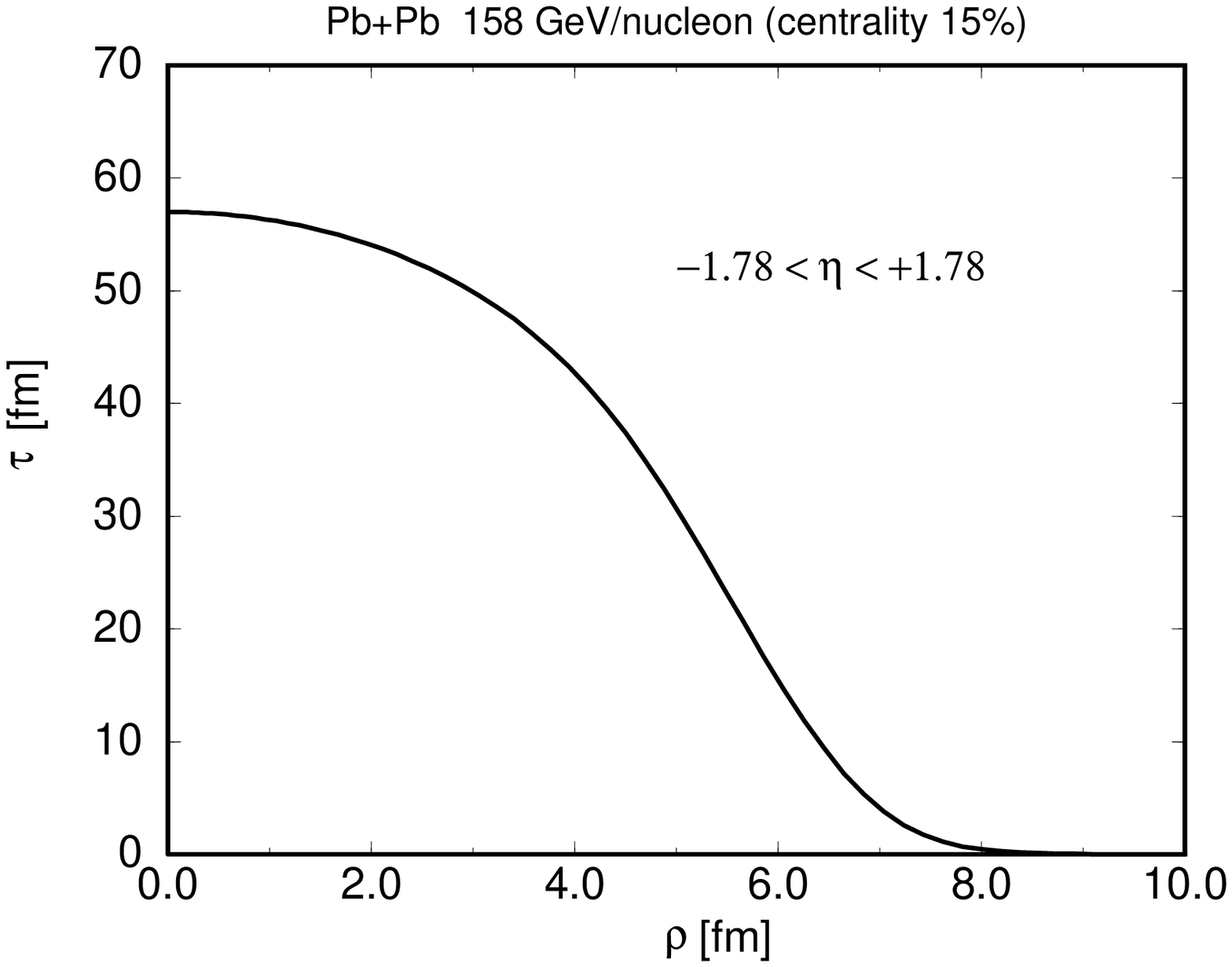}}}
\caption{}
\end{figure}
\begin{figure}[tbh]
{\vspace*{1.0cm} {\hspace*{+2.0cm} \epsfysize=9.5cm %
\epsfbox{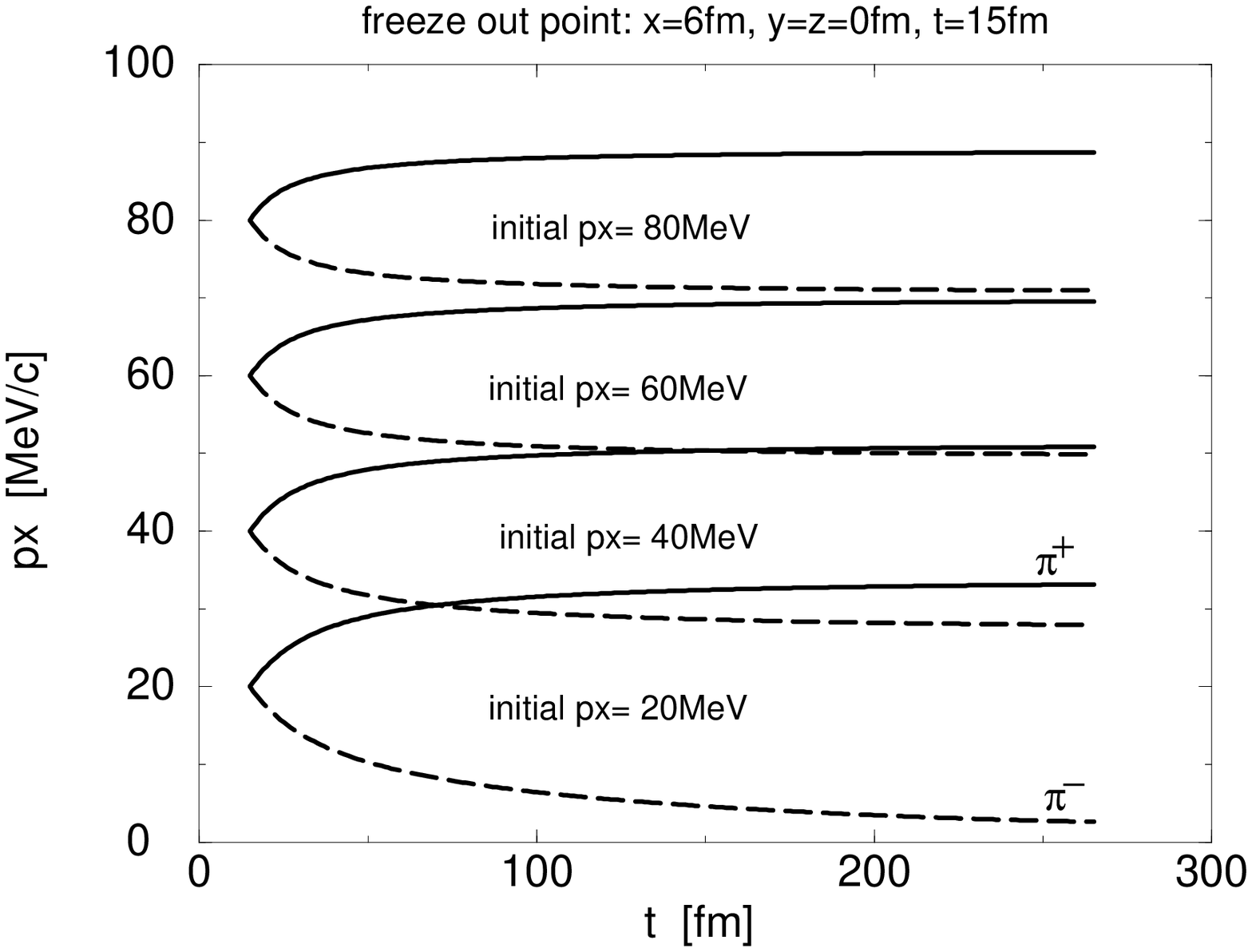}}}
\caption{}
\end{figure}
\begin{figure}[tbh]
{\vspace*{1.0cm} {\hspace*{+2.0cm} \epsfysize=9.5cm %
\epsfbox{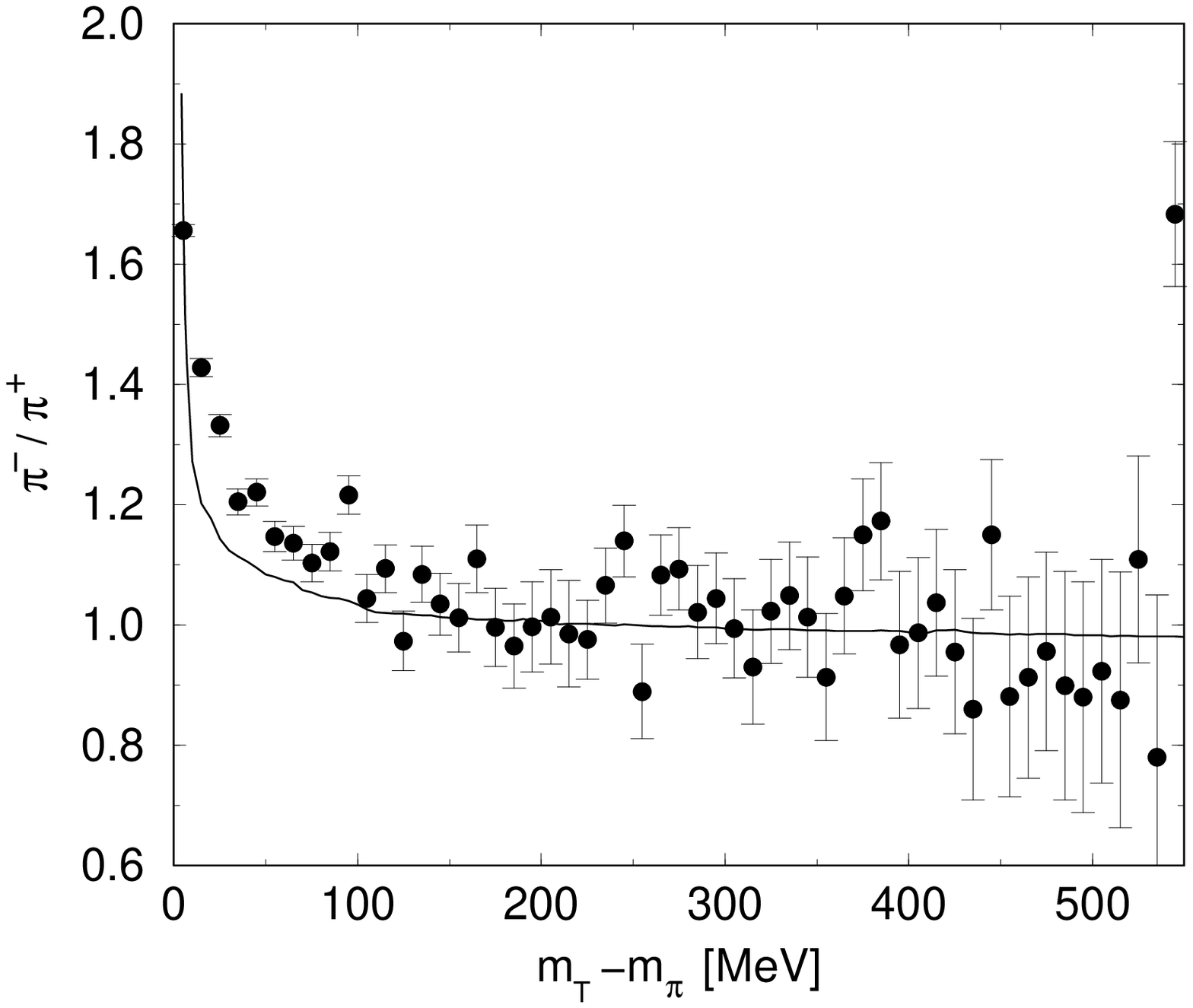}}}
\caption{}
\end{figure}
\begin{figure}[tbh]
{\vspace*{1.0cm} {\hspace*{+2.0cm} \epsfysize=9.5cm %
\epsfbox{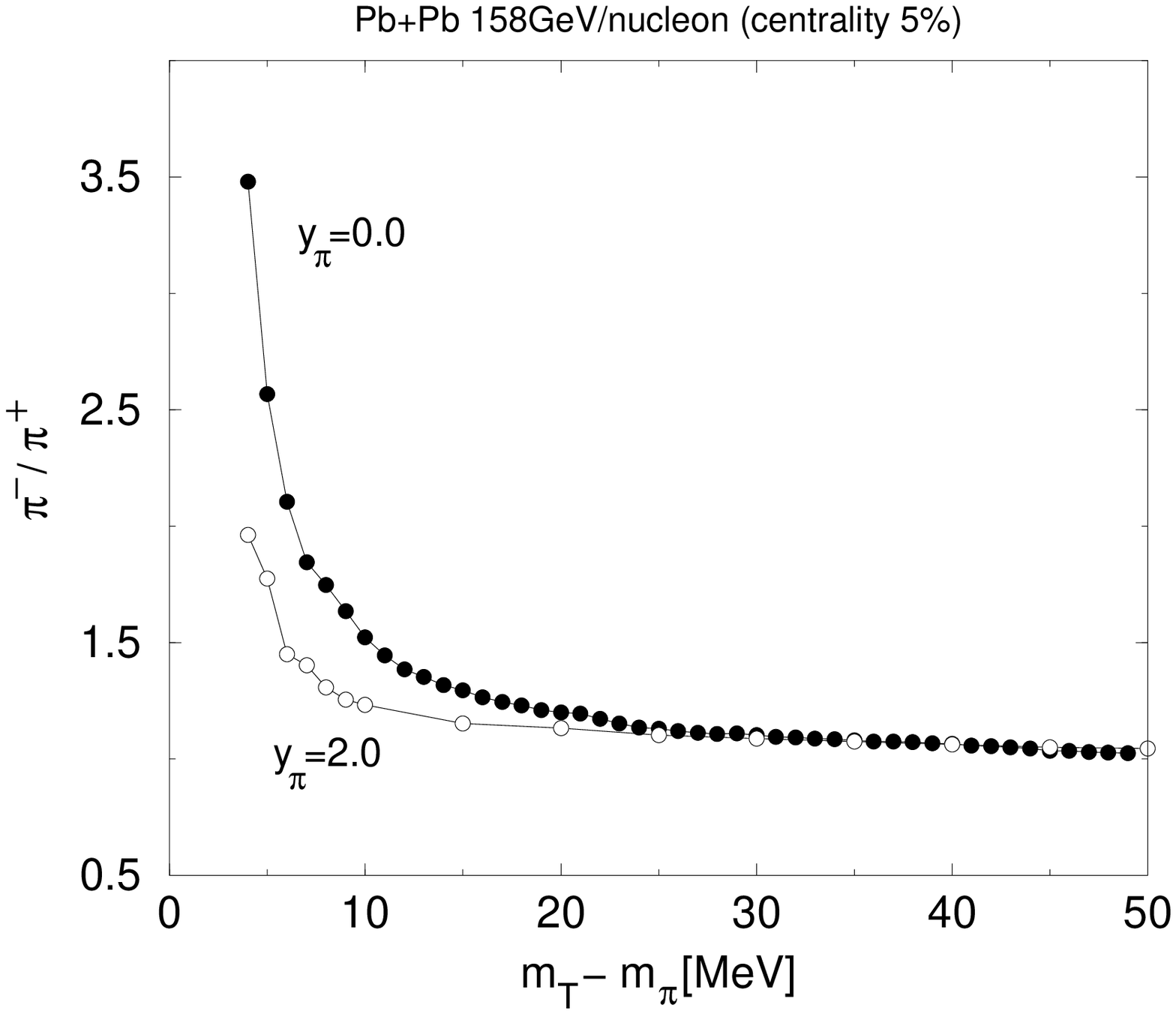}}}
\caption{}
\end{figure}
\begin{figure}[tbh]
{\vspace*{1.0cm} {\hspace*{+2.0cm} \epsfysize=9.5cm %
\epsfbox{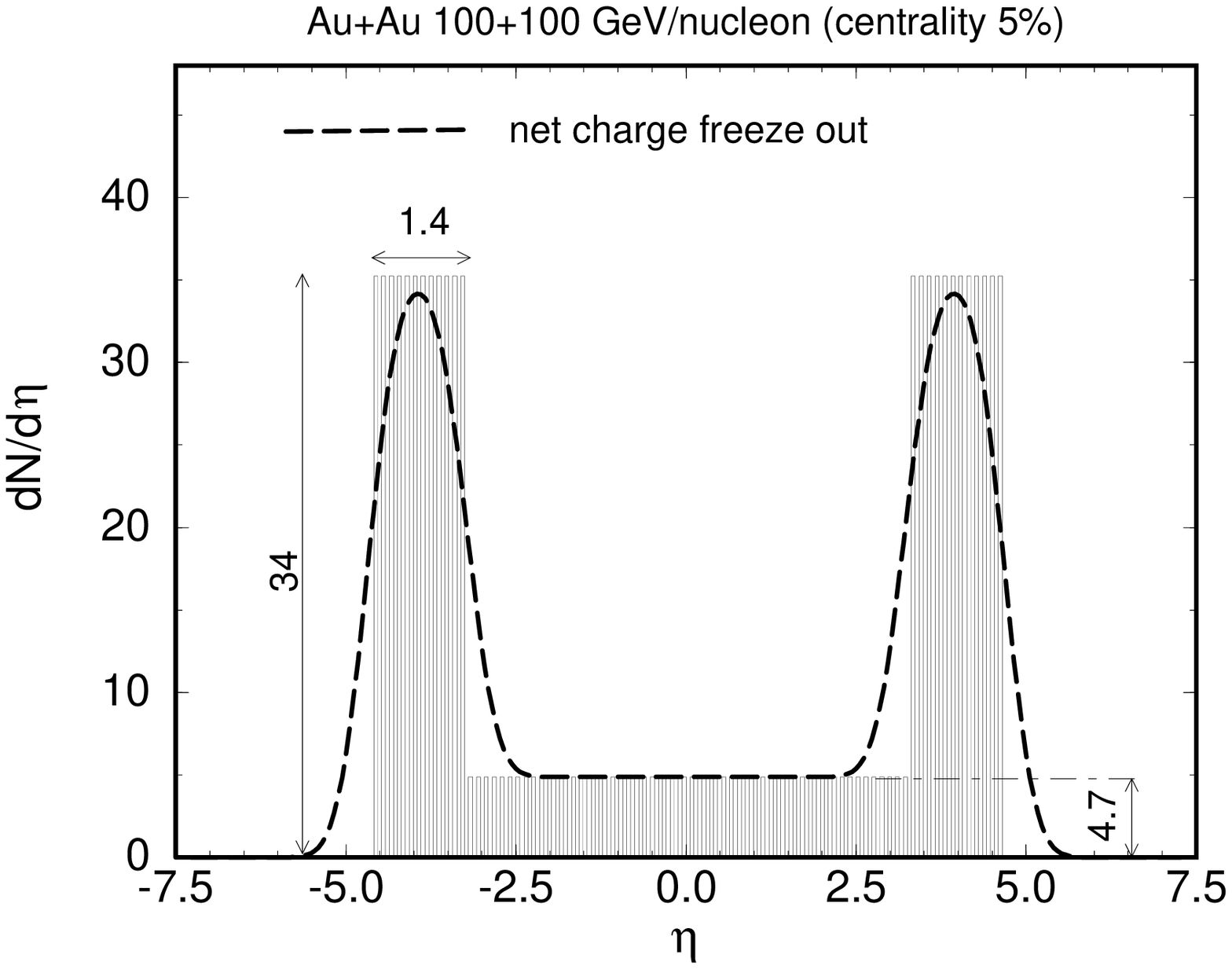}}}
\caption{}
\end{figure}
\begin{figure}[tbh]
{\vspace*{1.0cm} {\hspace*{+2.0cm} \epsfysize=9.5cm %
\epsfbox{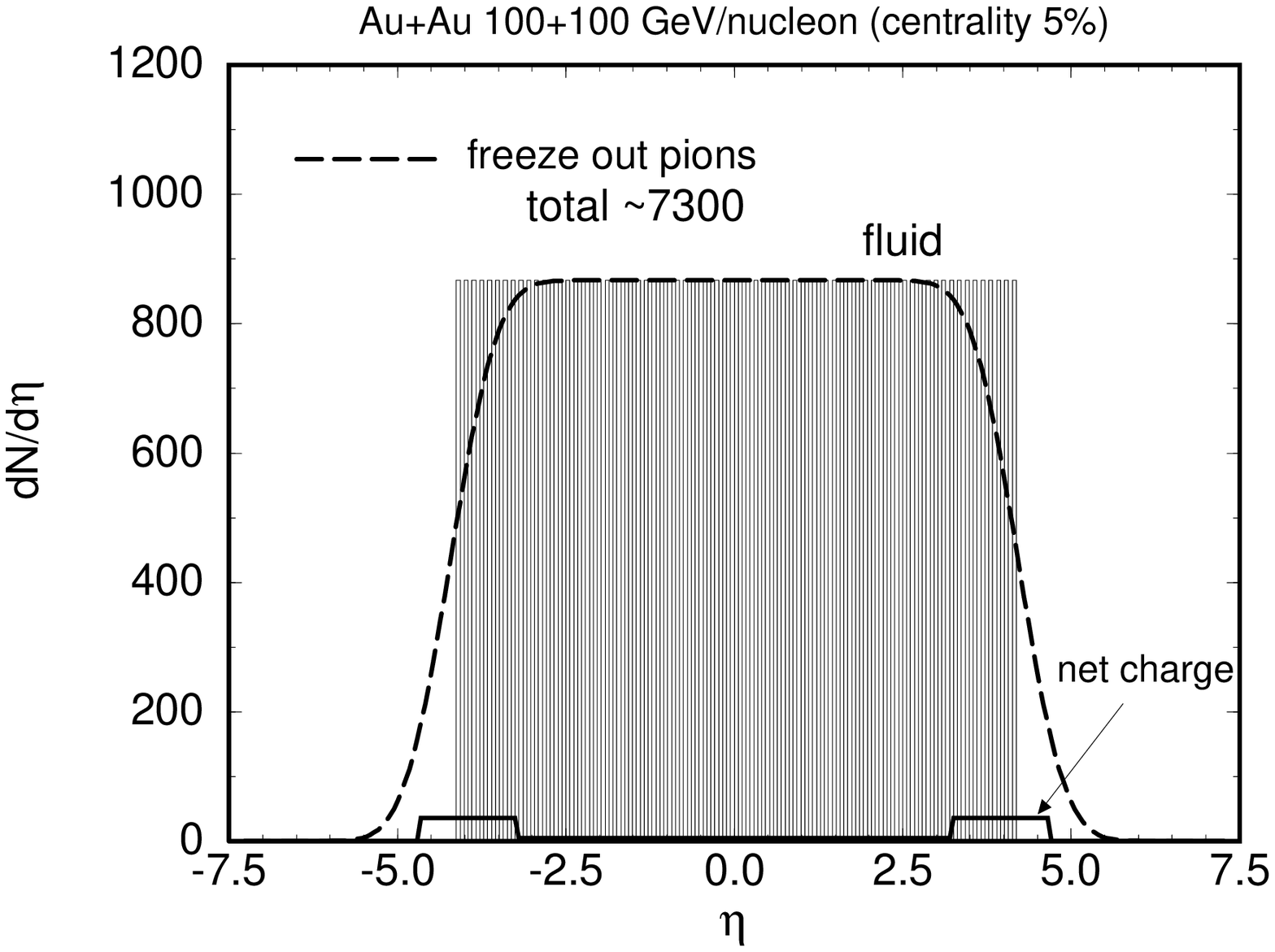}}}
\caption{}
\end{figure}
\begin{figure}[tbh]
{\vspace*{1.0cm} {\hspace*{+2.0cm} \epsfysize=9.5cm %
\epsfbox{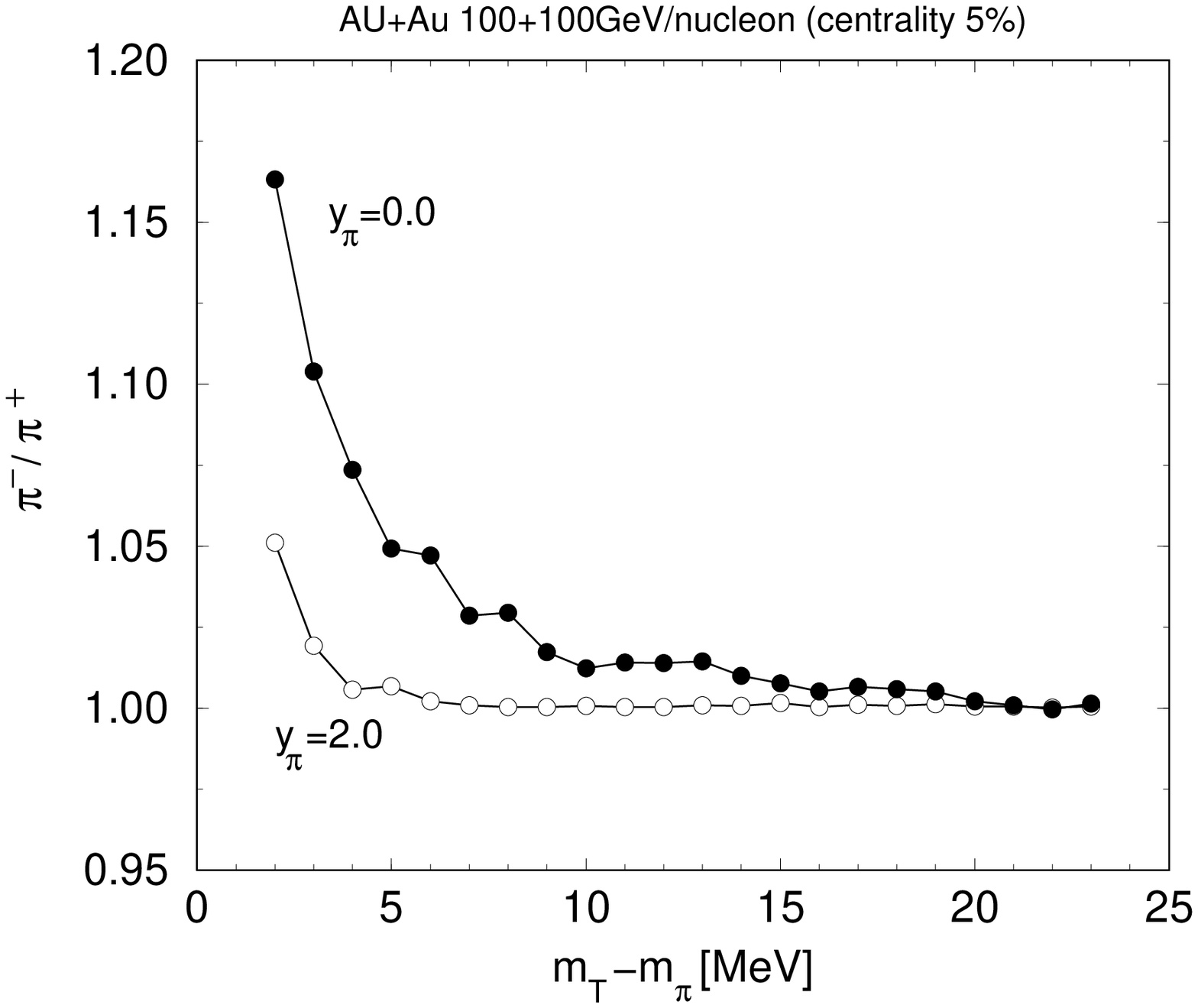}}}
\caption{}
\end{figure}
\end{document}